\documentclass[12pt,preprint]{aastex}
\begin{document}

\title{Proplyds and Massive Disks in the Orion Nebula Cluster 
Imaged with CARMA and SMA}

\author{J. A.  Eisner\altaffilmark{1,2}, R. L. Plambeck}
\affil{University of California at Berkeley \\ 
Department of Astronomy \\
601 Campbell Hall \\
Berkeley, CA 94720}
\email{jae@astro.berkeley.edu,plambeck@astro.berkeley.edu}

\author{John M. Carpenter, S. A. Corder}
\affil{California Institute of Technology \\ 
Department of Astronomy MC 105-24 \\
Pasadena, CA 91125}
\email{jmc@astro.caltech.edu,sac@astro.caltech.edu}

\and

\author{C. Qi, D. Wilner}
\affil{Harvard-Smithsonian Center for Astrophysics \\
60 Garden Street, Mail Stop 42 \\
Cambridge, MA 02138}
\email{cqi@cfa.harvard.edu,dwilner@cfa.harvard.edu}


\altaffiltext{1}{Miller Institute for Basic Research in Science, Berkeley, 
CA 94720}
\altaffiltext{2}{Steward Observatory, University of Arizona, Tucson, AZ 85721}

%

\begin{abstract}
We imaged a $2' \times 2'$ region of the Orion Nebula cluster in 1.3 mm
wavelength continuum emission with the recently commissioned 
Combined Array for
Research in Millimeter Astronomy (CARMA) and with the Submillimeter Array
(SMA)\footnote{The Submillimeter Array is a joint
project between the Smithsonian Astrophysical Observatory and the
Academia Sinica Institute of Astronomy and Astrophysics, and is
funded by the Smithsonian Institution and the Academia Sinica.}.  
Our mosaics include $\ga 250$
known near-IR cluster members, of which 36 are so-called ``proplyds''
that have been imaged previously with the Hubble Space Telescope.
We detected 40 sources in 1 mm continuum emission (one of which is
the BN Object), and several of them
are spatially resolved with our observations.  33 detected sources are
known near-IR cluster members, of which 11 are proplyds.  The 1 mm
emission from the majority of detected sources appears to trace
warm circumstellar dust.  However, for many of the proplyds, which are
located close to the Trapezium stars, the millimeter wavelength fluxes
are dominated by thermal free-free emission from hot, ionized gas.
Dust masses inferred for detected sources range from 0.01 to 0.5 M$_{\odot}$.
For the $\sim 225$ known near-IR cluster members not detected in our 1 mm 
observations, images toward the positions of near-IR sources 
were stacked to constrain the mean 1 mm flux 
of the ensemble.  The average flux is detected at the $\ga$ 4$\sigma$ 
confidence level, and implies an average disk mass of $\sim 0.001$ 
M$_{\odot}$, approximately an order of magnitude smaller than the 
minimum mass solar nebula.  Most stars in the ONC thus do not
appear to currently possess sufficient mass in small dust grains to form 
Jupiter-mass (or larger) planets.  Comparison with previous results
for younger and older regions indicates that massive disks evolve 
significantly on $\sim$Myr timescales.  We also show that the
percentage of stars in Orion surrounded by disks more massive than 
$\sim 0.01$ M$_{\odot}$ is substantially lower than in Taurus, indicating
that environment has an impact on the disk mass distribution.
Disks in Orion may be truncated through photoevaporation caused by the 
intense radiation field of the Trapezium stars, and we see marginal evidence
for such a scenario in the spatial distribution of massive disks
within the cluster.  Our data show no statistically significant
correlation between disk and stellar masses, although we see hints of
a higher percentage of massive disks around lower mass stars.
\end{abstract}


\section{Introduction}
The existence of protoplanetary disks around young stars is now firmly
established.  High resolution images from optical to radio wavelengths
have shown disk-like morphologies and Keplerian rotation profiles around
a number of young stars \citep[e.g.,][]{OW96,MO96,PADGETT+99,EISNER+04,KS95,
DUTREY+96,WILNER+00}.  Moreover, observations of near-IR emission from
young stars in excess of that expected from their 
stellar photospheres imply that most stars aged less than a few million years
possess inner circumstellar disks \citep[e.g.,][]{STROM+89,HLL01b}.

Protoplanetary disks are the birth-sites of planetary systems, and the 
ubiquity, properties, and lifetimes of disks constrain the
timescales and mechanisms of planet formation.  The mass distribution of 
protoplanetary disks is especially important since disk mass is related to the 
mass of planets that may potentially form.  For our own solar system,
the masses of planets and other bodies can be used to reconstruct a
minimum-mass solar nebula (MMSN) describing the 
amount of solar-composition material needed to build the solar system.  
Depending primarily on the core masses (and hence chemical compositions)
of Jupiter and Saturn (which are not known precisely), estimates of the
MMSN range from $\sim 0.01$--0.1 M$_{\odot}$ \citep{WEID+77}. Such disk masses 
are also required by planet formation models to build giant planets on
timescales shorter than inferred disk lifetimes
\citep[e.g.,][]{HAYASHI81,ALIBERT+05}.  The MMSN is thus an
informative benchmark against which to gauge the potential of disks around 
other stars to form solar systems like our own.

A widely used method \citep[e.g.,][]{BECKWITH+90}
for measuring disk masses is to observe emission 
from optically thin dust, and then use assumed dust grain properties
to convert observed fluxes into dust masses.  An assumed gas-to-dust ratio
is then used to estimate the total (gas+dust) circumstellar mass.
At short wavelengths ($\lambda \la 10$ $\mu$m), the dust in protoplanetary
disks is optically thick even for masses $<10^{-6}$ M$_{\odot}$.  
Observations at sub-mm and mm wavelengths are necessary to measure 
optically thin dust emission, and hence to determine the total mass
of dust in the disk.  


Several investigators have carried out 
comprehensive single-dish mm and sub-mm continuum surveys toward
regions of star formation comprising loose aggregates of stars:
Taurus \citep{BECKWITH+90,OB95,MA01,AW05}, $\rho$ Ophiuchi 
\citep{AM94,NUERNBERGER+98,MAN98,AW07}, 
Lupus \citep{NCZ97}, Chamaeleon I \citep{HENNING+93}, Serpens
\citep{TS98}, and MBM 12 \citep{ITOH+03,HOGERHEIJDE+02}.  
In Taurus, 22\% of stars surveyed appear to possess disks more massive than
$\sim 0.01$ M$_{\odot}$, and the median disk mass is 
$5 \times 10^{-3}$ M$_{\odot}$ \citep{AW05}.  The fraction of massive 
disks\footnote{Here and throughout the text, ``massive disks'' refer
to disks with mass comparable to or greater than 0.01 M$_{\odot}$, the
lower range of estimates for the minimum-mass solar nebula.} and
the median disk mass is comparable in $\rho$ Ophiuchi \citep{AM94,AW07}.

Low-density star forming regions are not the typical birth-sites
of stars; rather, most stars form in rich clusters like the Orion Nebula
\citep{LADA+91,LSM93,CARPENTER00,LL03}.  Isotopic abundances in 
our solar system suggest that it, too, may have formed in 
a dense, Orion-like environment \citep[e.g.,][]{HD05,WG07}. 
Expanding millimeter continuum surveys to include rich clusters allows the 
determination of the frequency and evolution of massive disks in
typical star (and planet) formation environments.  The high stellar density
in rich clusters also allows the assembly of good statistics since many 
disks can be mapped at once.  

The main challenge to observing rich clusters at (sub-)mm 
wavelengths is that very 
high angular resolution is required to resolve individual sources
and to distinguish compact disk emission from the more extended emission
of the molecular clouds in which young clusters are typically embedded.  
Single-aperture mm-wavelength telescopes lack sufficient angular resolution, 
and to date, only three rich clusters have been observed with mm-wavelength 
interferometers: the Orion Nebula cluster \citep{MLL95,BALLY+98,WAW05,EC06}, 
IC 348 \citep{CARPENTER02}, and NGC 2024 \citep{EC03}. 

These interferometric surveys of rich clusters have detected very few disks
with $\ga 0.01$--0.1 M$_{\odot}$ of material, in large part because of
limited sensitivity and areal coverage.  The most recent observations of Orion
detected emission from several massive ($\ga 0.01$ M$_{\odot}$) disks 
\citep{WAW05,EC06}, while upper limits from other surveys range from 
$\sim 0.025$--0.17 M$_{\odot}$ \citep{MLL95,BALLY+98}.
Considering as ensembles the large numbers ($\ga 100$) of young stars included 
in the cluster surveys allowed estimates of 
mean disk masses of $\sim 0.002$, 0.005, and 0.005 M$_\odot$ in 
IC 348, NGC 2024, and the ONC, respectively
\citep{CARPENTER00,EC03,EC06}. Thus, it appears that many stars
aged $\la 1$ Myr still possess massive circumstellar disks, although
more sensitive observations are necessary to detect directly large numbers
of massive disks at a range of ages, and thereby constrain the mass 
distribution and evolutionary timescales.

Here we present a new 1.3 mm wavelength interferometric survey of
the Orion Nebula cluster (ONC), a young, embedded
stellar cluster that includes the bright, massive Trapezium stars.
Our observations make use of the Submillimeter Array (SMA) and the recently 
commissioned Combined Array for Research in Millimeter Astronomy (CARMA).
The combination of data from these two instruments yields a map of
the mm-wavelength continuum emission in Orion with unprecedented 
sensitivity, angular resolution, and image fidelity.

The Trapezium region contains hundreds of stars within a several
arcminute radius, and pre-main-sequence
evolutionary models \citep[e.g.,][]{DM94} fitted to spectroscopic
and/or photometric data indicate that most stars are less than 
approximately one million years old \citep[e.g.,][]{PROSSER+94,HILLENBRAND97}.
The standard deviation in the distribution of 
inferred stellar ages is $\la 1$ Myr \citep{HILLENBRAND97}.
Our observations thus provide a snapshot of millimeter emission around
a large number of roughly coeval young stars.  

With the large number of stars in the ONC, we can also investigate the
correlation of disk properties with stellar and/or environmental
properties.  Previous investigations of near-IR excess emission
showed the inner disk fraction for stars in Orion
to be largely independent of stellar age and mass, although there are
indications of a paucity of disks around very massive stars
\citep{HILLENBRAND+98,LADA+00}.  In addition, the inner disk fraction may
decrease at larger cluster radii \citep{HILLENBRAND+98}.
The millimeter observations presented here enable investigation of
how such stellar and environmental properties correlate with disk mass.

We adopt a distance to the ONC of 400 pc, based on
recent trigonometric parallax measurements of several stars 
\citep{SANDSTROM+07,MENTEN+07} and orbital fitting for a spectroscopic
binary \citep{KRAUS+07}.  This is substantially lower than the value
of 480 pc computed based on statistical parallax of water maser spots
\citep{GENZEL+81}, which was adopted in previous studies of the ONC
\citep[e.g.,][]{EC06}.  We discuss below the importance of this revised
distance on our results.



\section{Observations and Data Reduction \label{sec:obs}}

\subsection{CARMA Observations and Calibration \label{sec:carma_obs}}
We mosaicked a $2' \times 2'$ region toward the
ONC in $\lambda$1.3 mm continuum with CARMA between October and December, 2007.
CARMA consists of six 10-m antennas and nine 6-m antennas situated at 
2200 meters elevation at Cedar Flat in the Inyo Mountains of California.  
With a total of 15 antennas, CARMA provides 105 baselines, enabling excellent 
coverage of the uv plane and hence high image fidelity.  Two different array
configurations (`C' and `B') were used to obtain antenna separations
ranging from 30 to 946 meters.

Continuum data were recorded in six $\sim 500$ MHz bands covering the
frequency ranges 221.75--223.25 GHz and 227.25--228.75 GHz from the
receivers' lower and upper sidebands, respectively.  Each band
consists of 15 channels.  Spectral line emission in Orion is mostly resolved
out by these observations; that is, across most of the mapped region
it is detected only weakly because
it is spatially extended relative to the interferometer fringe
spacings.  The spectral lines most visible on the shortest baselines,
mostly toward the BN/KL region in the northwest corner of the mosaic, are the
11(1,11)-10(0,10) transition of SO$_2$ at 221.965 GHz and the 25-24
transition of HC$_3$N at 227.402 GHz.  Even these lines are almost
completely resolved out for projected antenna spacings $> 70$
k$\lambda$ (used to generate the final maps; see \S \ref{sec:mapping}). 
We therefore assume our bands are effectively line-free.

\begin{figure}
\plotone{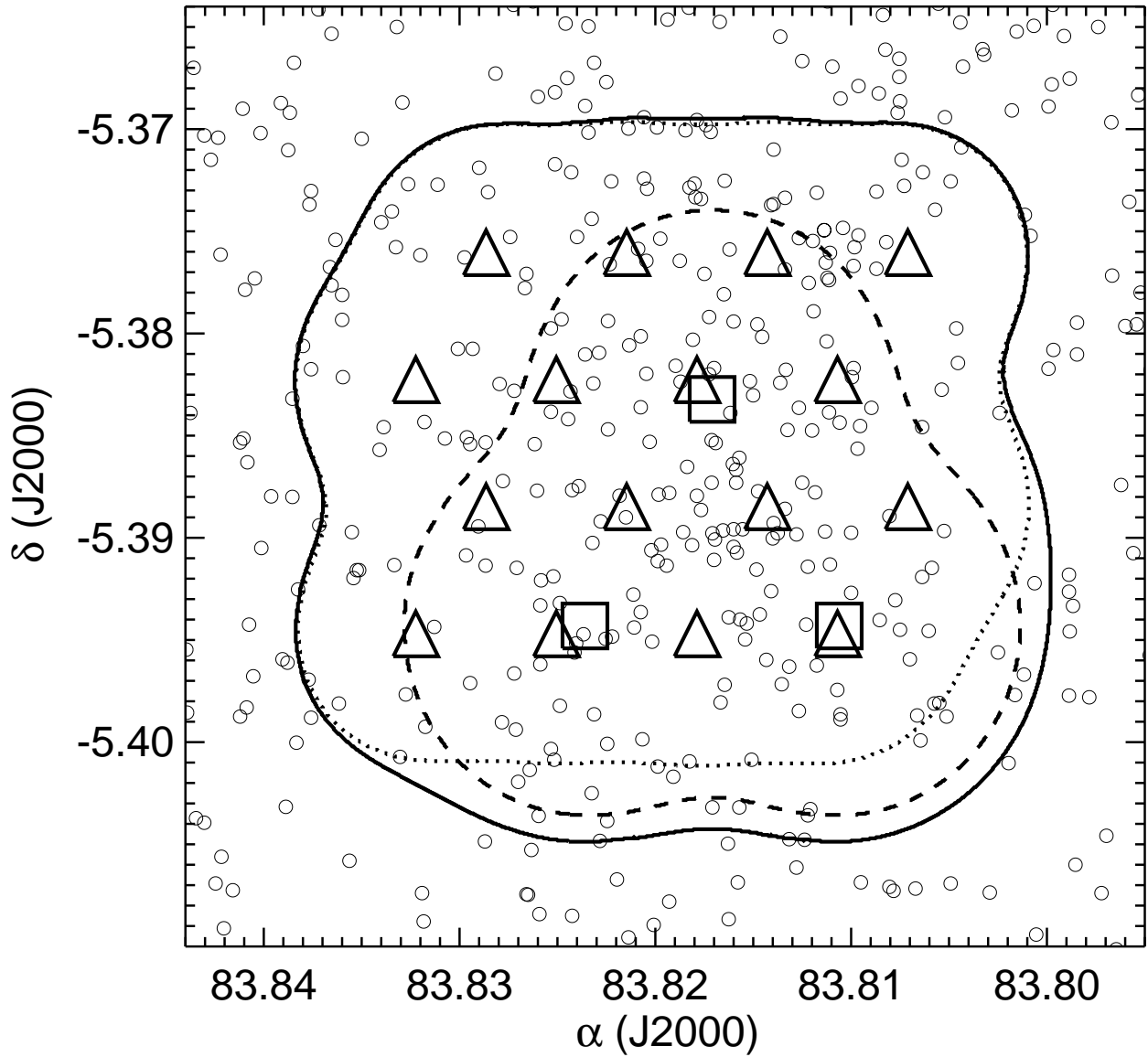}
\caption[Pointing positions for the ONC mosaics]
{Pointing positions for the CARMA mosaic (triangle symbols) and
the SMA mosaic (square symbols) plotted over the positions of $K$-band sources 
in the ONC (open circles).  These $K$-band source positions are drawn
from \citet{HC00}, and have been registered to the 2MASS astrometric grid.
The unit gain contour of the CARMA mosaic (dotted curve), the SMA mosaic
(dashed curve), and the mosaic produced from the combined SMA+CARMA dataset
(solid curve) are also indicated.
\label{fig:pointings}}
\end{figure}

The mosaic consists of 16 pointing centers (Figure \ref{fig:pointings}), 
separated by
$26''$.  This separation is comparable to the FWHM beamwidth of the 10-m
antennas, but $\sqrt{3}$ smaller than the FWHM beamwidth of the 6-m
antennas.  A 2-D mosaic is Nyquist sampled if pointings are separated
by $\le $FWHM/$\sqrt{3}$.  Thus the CARMA mosaic is Nyquist sampled only for
the 6-m dishes.  However, Nyquist sampling is not crucial since we 
are interested only in compact sources, rather than extended emission.
Simulated CARMA mosaics of a synthetic star field showed that $26''$
spacings provided the best balance of sensitivity and areal
coverage to maximize the number of sources detected.

Each night we interleaved 16-minute observations of the ONC mosaic, 
with 1 minute integration time on each pointing center, with 3-minute
observations of the phase and amplitude calibrator, J0530+135.
Observing the mosaic in its entirety every 20 minutes and repeating
this multiple times during the night ensures a high quality
synthesized beam and equal sensitivity for each pointing center.  
The total integration time for the maps was 34 minutes per pointing
center.  

Telescope pointing was checked and updated on 20 
minute intervals using optical counterparts very near to the source.
These objects were observed with optical cameras mounted on the dishes,
and radio--optical offsets were calibrated periodically.
Gain stability, especially in 1 mm observations like those presented here, is 
found to be enhanced through this method (Corder, Carpenter 
\& Wright, in prep). 

We measured a 1.3 mm flux density of $3.0 \pm 0.3$ Jy for J0530+135
using Uranus as a primary flux calibrator, based on observations
during several nights in October when both sources were observed.
J0530+135 was also used to calibrate the passband.  
All calibrations
for these data were performed with the MIRIAD package \citep{STW95}.

\subsection{SMA Observations and Calibration \label{sec:sma_obs}}
We mosaicked a triangular region approximately $1\rlap{.}'7$ on a side,
consisting of three pointings, with the SMA between September, 2005, and
February, 2006
(Figure \ref{fig:pointings}).  
The SMA consists of eight 6-m dishes near the summit of Mauna Kea in 
Hawaii.  However, for our observations only seven antennas were available,
providing 21 baselines between 10 and 220 m. 
The $uv$ coverage for
our SMA observations is substantially sparser than for our CARMA observations,
and thus the image fidelity is worse (which means, for example, that
strong emission is scattered more strongly into other regions of the map).  
However, the high altitude of
the SMA enables very low opacity observing conditions.  The low opacities,
combined with increased observing time per pointing, lead to substantially 
better sensitivity relative to the CARMA observations.

Double sideband receivers were tuned to a local oscillator (LO) frequency of
225.333 GHz. The SMA digital correlator is configured with 24 partially
overlapping bands of 104 MHz width in each sideband. Each sideband
provides 2 GHz of bandwidth, centered $\pm 5$ GHz away from the LO frequency.
The double-sideband
(DSB) system temperatures were between 80 and 200 K.

As for the CARMA mosaic, we
observed the mosaic multiple times throughout the night, obtaining
equal integration times for each pointing position.
The pointings in the mosaic were separated by $\sim 44''$, the approximate 
FWHM for the SMA dishes at this observing wavelength.  As for the CARMA
map, the SMA mosaic is larger-than-Nyquist sampled, which provides
enhanced areal coverage compared to a Nyquist-sampled mosaic.

We used J0423-013 and 3C120 as gain calibrators for these tracks, with
flux densities of 1.4 and 1.0 Jy, respectively, derived using Uranus
as a primary flux calibrator.  We estimate that the absolute flux scale is
uncertain by $\sim 10\%$.  We calibrated the passband
using the quasar 3C454.3, and Uranus where available.  
Because the CO(2-1) transitions are present in
the observing window, we edited out the parts of the band with 
strong lines and
generated a line-free continuum channel.  All calibrations were performed
using the SMA adaptation of the IDL-based data reduction package MIR
developed at Caltech;
calibrated data were then converted into MIRIAD format for further processing.

\subsection{Mapping \label{sec:mapping}}
We made mosaics of our CARMA and SMA datasets individually, and after combining
the two datasets in the $uv$ plane.  For the individual and combined
datasets, we mosaicked the individual pointings 
into a single image, weighting the data by system temperature and by $uv$
distance (with a ``robust'' parameter of 0.5), then de-convolved and 
CLEANed (all using MIRIAD). The angular resolution afforded by 
the longest baseline data in our maps (from the CARMA B-array) 
is $\sim 0\rlap{.}''3$.   Our mosaics have $0\rlap{.}''1$ 
pixels, which ensures adequate sampling of individual resolution elements.

Since we are primarily interested in compact disk emission, we 
eliminated $uv$ spacings shorter than 70 k$\lambda$ (i.e., projected
baselines shorter than 93 m) in order to reduce 
contamination from bright extended emission.  The eliminated spacings
correspond to size scales larger than $\sim 3''$.  The cutoff  value was chosen
to minimize the RMS background noise in the CLEANed images; we measured
the RMS for $uv$ cutoff radii of 50, 60, 70, and 80 k$\lambda$, and found
the 70 k$\lambda$ cutoff to be optimal. 

\begin{figure}
\plotone{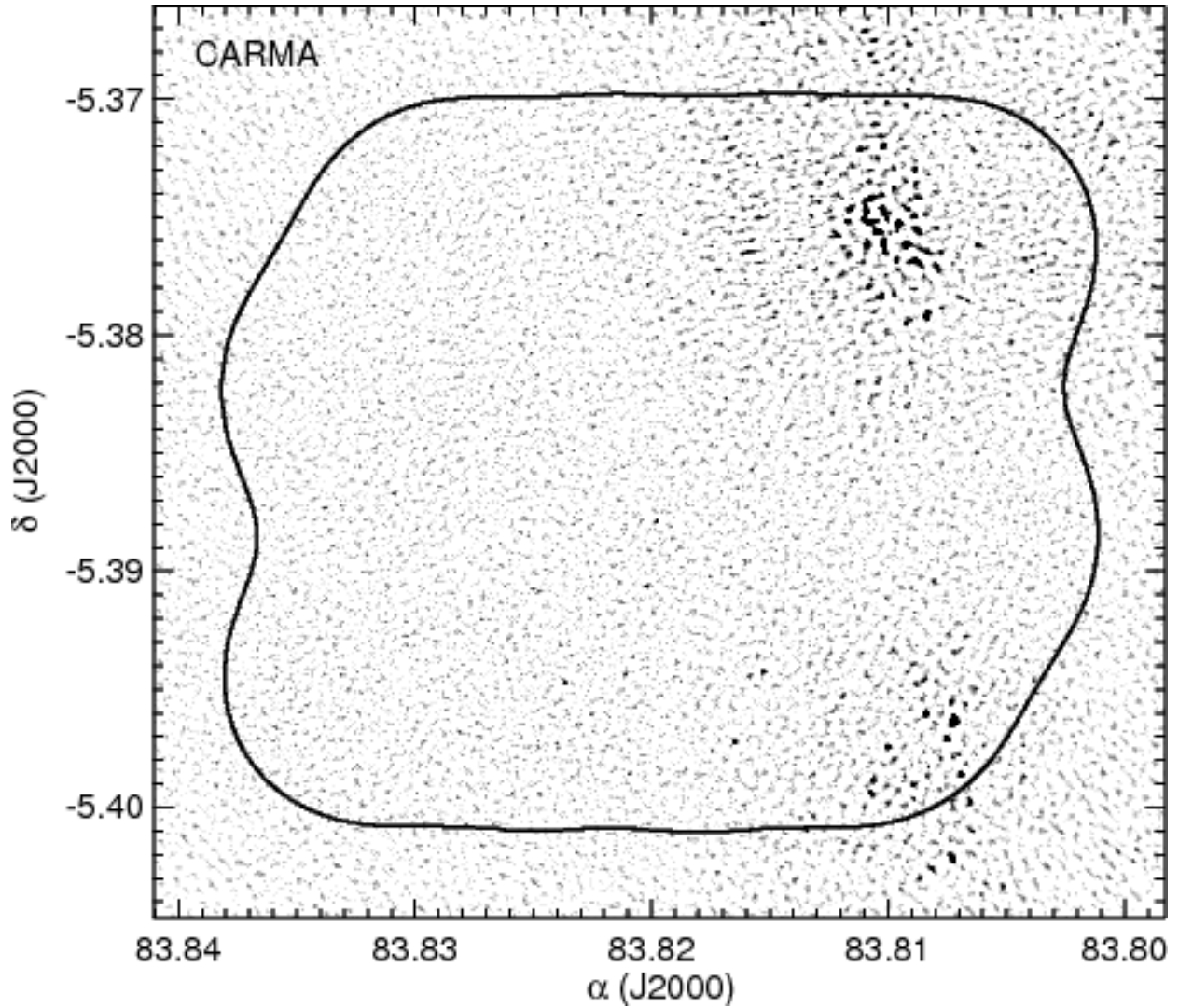}
\caption{The Orion Nebula cluster, imaged in $\lambda$1.3 mm
continuum with CARMA.  Only data
observed on long baselines ($r_{uv}>70$ k$\lambda$) were
used to create this image.  The unit gain region of the mosaic 
is indicated by a solid curve. The synthesized beam
has FWHM of $0\rlap{.}''61 \times 0\rlap{.}''52$ at a PA of $70^{\circ}$.
The RMS noise varies within the unit gain contour from 2.3 to 23 mJy 
(1$\sigma$), with a mean RMS of 4.6 mJy.   Partially resolved extended
emission in the BN/KL and OMC1-S regions increases the noise level in
the upper and lower right corners of the image.  The map has been divided
by the theoretical RMS, to visually down-weight the noisier edges.
\label{fig:mos}}
\end{figure}

\begin{figure}
\plotone{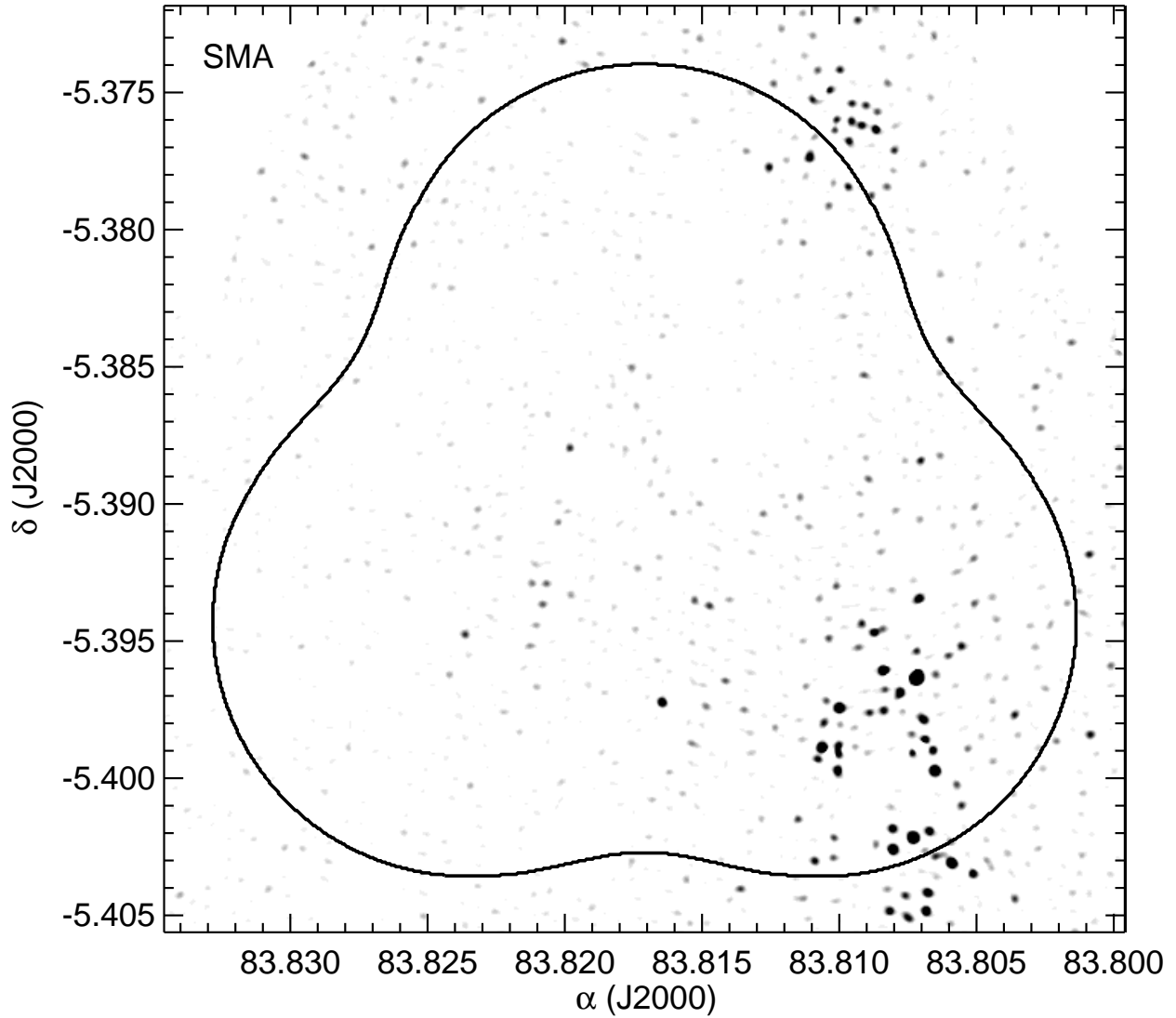}
\caption{The Orion Nebula cluster, imaged in $\lambda$1.3 mm
continuum with the SMA.  Only data
observed on long baselines ($r_{uv}>70$ k$\lambda$) were
used to create the image.  The unit gain region of the mosaic 
is indicated by a solid curve.  The synthesized beam has a roughly circularly 
symmetric core with a FWHM of $0\rlap{.}''98$.  The RMS within the unit gain 
contour varies from 0.8 to 28 mJy, with a mean value of 2.7 mJy.  The BN/KL 
and OMC1-S regions produce the strong emission visible in the upper and lower 
right corners of the map.
The map has been divided
by the theoretical RMS, to visually down-weight the noisier edges.
\label{fig:sma}}
\end{figure}

\begin{figure}
\plotone{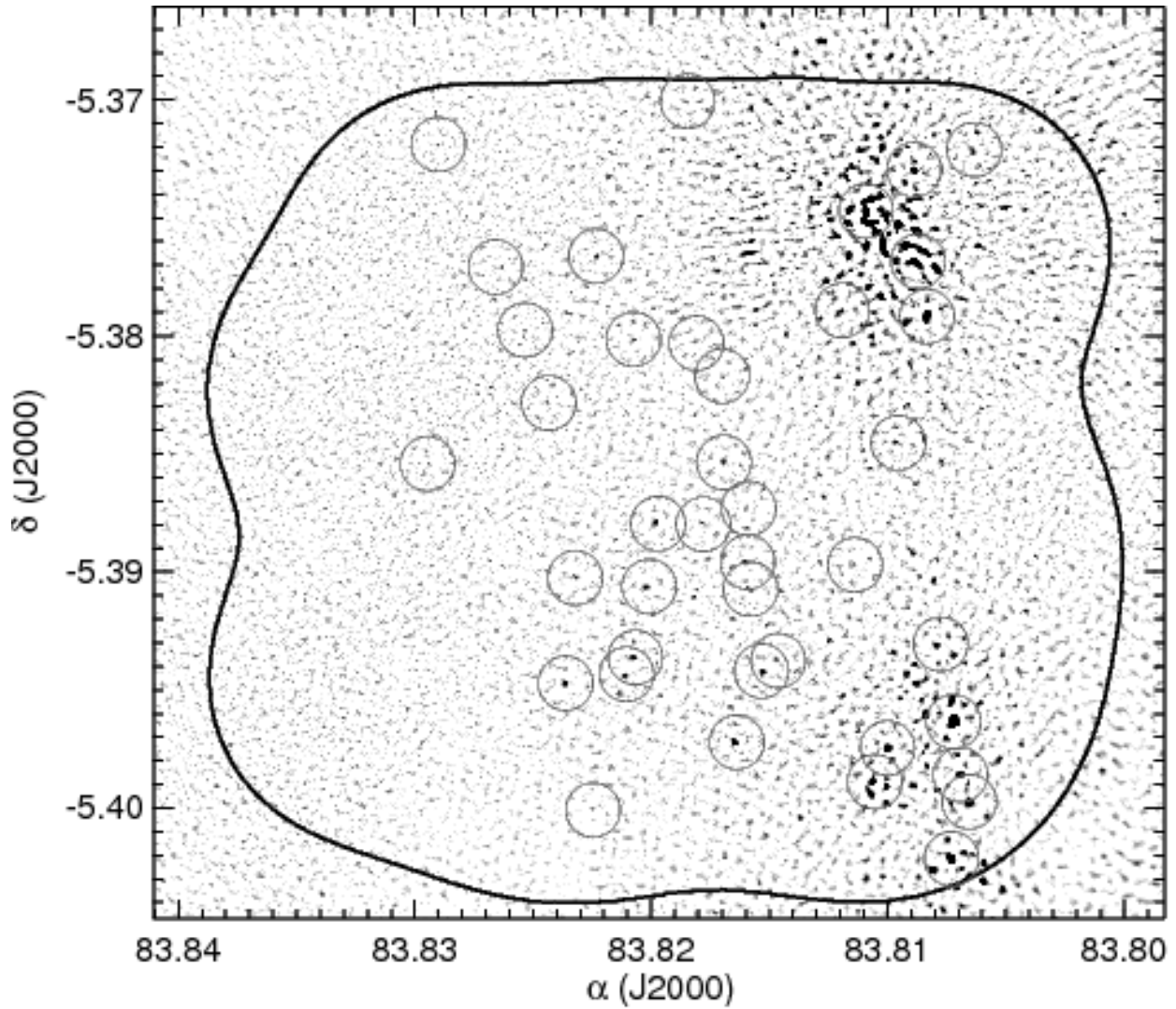}
\caption[$\lambda$1 mm continuum mosaic of the ONC]
{The Orion Nebula cluster, imaged in $\lambda$1.3 mm
continuum with CARMA and the SMA (greyscale).  CARMA and SMA
data were combined in the $uv$ plane, and only data
observed on long baselines ($r_{uv}>70$ k$\lambda$) were
used to create this image.  The angular resolution is 
$\sim 0\rlap{.}''7 \times 0\rlap{.}''6$.  The unit gain region of the mosaic 
encompasses a $2' \times 2'$ area, as indicated by the solid contour, 
and the RMS residuals (1$\sigma$) within this region vary from
1.8 mJy in regions devoid of bright emission to $\ga 10$ mJy in the
crowded regions toward the right of the map.
Millimeter detections above the 3$\sigma$ level coincident with 
near-IR cluster members and proplyds, and sources without counterparts
detected above the 5$\sigma$ level (\S \ref{sec:thresh}),  
are indicated by gray circles.
The map has been divided
by the theoretical RMS, to visually down-weight the noisier edges.
\label{fig:map_combo}}
\end{figure}

Mosaics produced from our robust-weighted data with $r_{uv} > 70$ k$\lambda$
are shown in Figures \ref{fig:mos}--\ref{fig:map_combo}.  
In the Figures we have divided by the theoretical sensitivity at each location 
in the image, in order to visually down-weight the noisier edges of
the mosaic (where there are fewer overlapping pointings); we do not
divide by the sensitivity in the analysis presented below.  
We note that even within the uniform (theoretical) 
sensitivity region of the mosaic,
the RMS varies substantially because of emission scattered from bright compact 
and extended sources in the BN/KL and OMC1-S regions located in the upper and
lower right quadrants of the maps.

For the SMA mosaic, the unit gain  region (within which the theoretical
sensitivity does not vary substantially) encompasses a roughly triangular
region covering $\sim 2$ square arcminutes.  The RMS of pixels
within ``clean''  regions of the unit gain contour (i.e., away from the
crowded BN/KL and OMC1-S regions) is 0.8 mJy.  
The unit gain region of the CARMA mosaic  
encompasses a $2' \times 2'$ area, with an RMS noise level (again, in
clean regions of the map) of 2.3 mJy.
For the combined map, which will be used for the bulk of our analysis,
the unit gain region is slightly larger than for the CARMA-only mosaic, and
the RMS noise level is 1.8 mJy.  The synthesized beam has dimensions (FWHM) of 
$0\rlap{.}''69 \times 0\rlap{.}''60$ at a position angle of 72$^{\circ}$.   

\section{Analysis and Results \label{sec:analysis}}

\subsection{Detection Thresholds \label{sec:thresh}}
Because the map contains a large number of pixels, we must employ a fairly
high detection threshold to avoid random noise spikes if we search the image
blindly.  The mosaic area is approximately $35,000$ synthesized beams.  With 
this number of independent pixels, one expects $> 1$ noise spike above 
the 4$\sigma$ level (assuming Gaussian noise).
We therefore use a 5$\sigma$ detection limit, at which level $\ll 1$ pixels are
expected to show noise spikes.  Because the noise varies greatly across the 
map, we calculate $\sigma$ locally in small sub-regions of the image.

Specifically, a ``local'' $\sigma$ is computed 
in $10'' \times 10''$ ($100 \times 100$ pixels) boxes around each pixel
in the mosaic.  For our detection thresholds to be meaningful, the 
noise must be well-characterized.  However, poorly sampled extended emission
leads to excess noise in the BN/KL and OMC1-S regions.   Moreover, the
noise increases toward the edges of the unit gain region because there
are fewer overlapping mosaic pointings there.
Detections in these areas should be treated with some caution.

As a test of our detection threshold, we searched the maps
for false detections below the $-5\sigma$ level.  None were detected,
confirming that 5$\sigma$ is a reasonable detection limit.  
In contrast, 28 sources were seen below the $-4\sigma$ level 
(most of them toward the edges of the mosaic or in the BN/KL and OMC1-S 
regions), demonstrating that 4$\sigma$ is not a sufficiently stringent 
detection threshold (and that the noise across our mosaic is not
always Gaussian).

Instead of blindly searching for detections, we can also use our prior 
knowledge of the locations of near-IR cluster members and search only these
positions.  For these $\sim 250$
pre-determined positions, $\sim 0.3$ sources are expected to show
emission above the 3$\sigma$ level from Gaussian noise.
We can therefore try a 3$\sigma$ detection threshold, 
where $\sigma$ is the noise determined locally 
(as above) in $10'' \times  10''$ sub-regions
centered on individual cluster member positions.  Although the
noise in the mosaic is not always Gaussian, this 3$\sigma$ threshold
appears reasonable: none of the near-IR source positions 
were detected below the $-3\sigma$ level.

Sources with 1 mm continuum emission at the $>3 \sigma$ level in our maps
are deemed to coincide with near-IR cluster members if the mm peaks
and near-IR source positions lie within $0\rlap{.}''4$ of each other.
The estimated relative positional accuracy of $0\rlap{.}''4$ is the
quadrature sum of uncertainties from centroiding the mm images 
($\sim 0.5 \theta_{\rm beam}/$signal-to-noise $\approx 0\rlap{.}''1$),
uncertainties in the absolute astrometry due to baseline errors
($\sim 0\rlap{.}''2$), and
uncertainties in the near-IR source positions ($\sim 0\rlap{.}''3$).  

\begin{figure}
\plotone{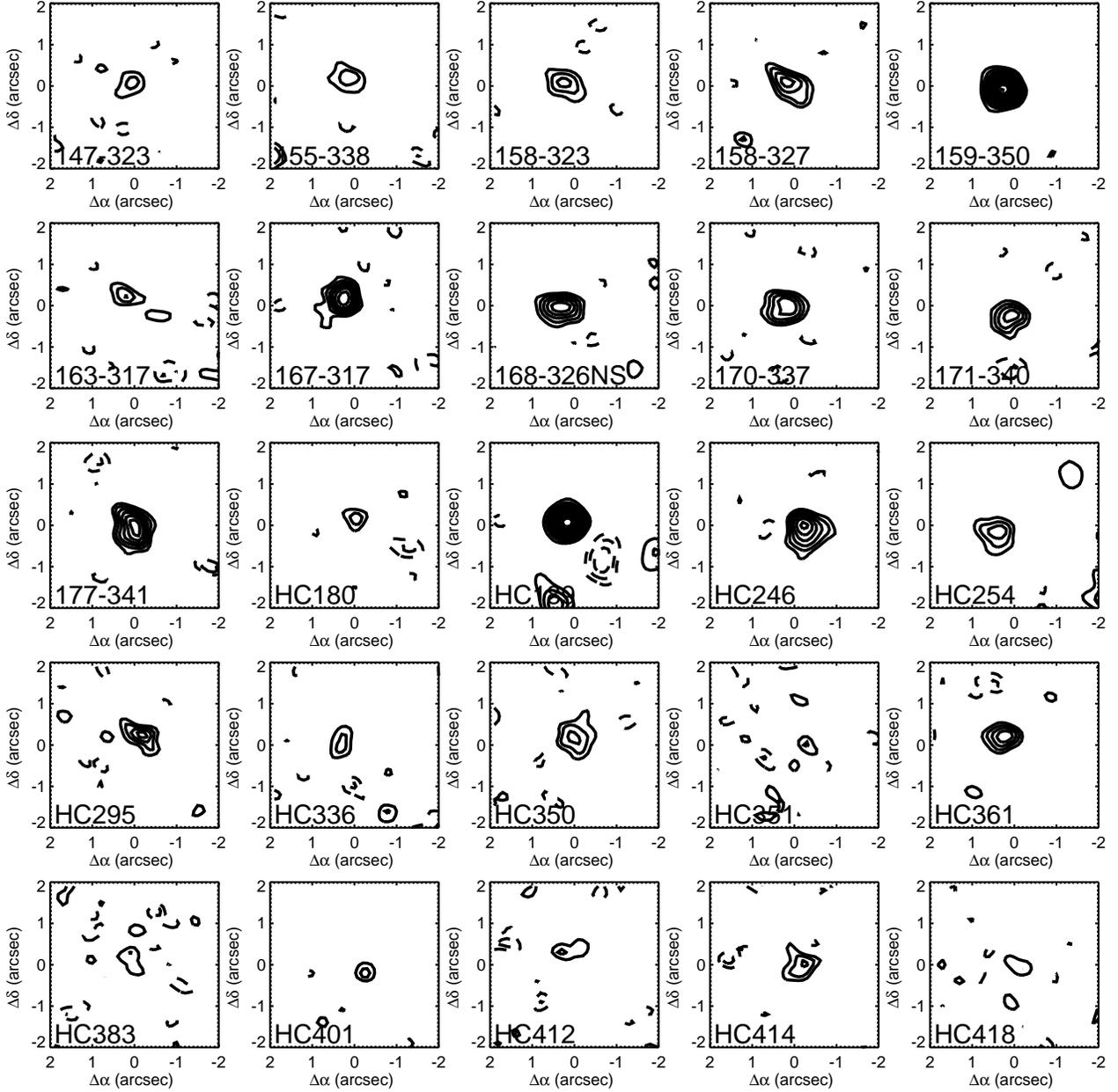}
\caption{Contour images of sources detected in 1 mm continuum emission.
Contour increments are 1$\sigma$, beginning at $\pm 2 \sigma$, where 
$\sigma$ is determined locally for each object (see \S \ref{sec:thresh}). 
Solid contours represent positive emission, and dotted contours trace negative 
features.  The BN object is excluded from these plots, as it is
discussed in more detail in a later paper.
For sources detected at infrared wavelengths, individual images are
centered on the previously measured near-IR coordinates.  For the MM
sources, detected only at $\ga 1$ mm wavelengths, images are
centered on the peak fluxes.
\label{fig:detections}}
\end{figure}

We detected 19 sources within the unit gain contour of
our mosaic above the 5$\sigma$ level (Table 
\ref{tab:detections}).  12 of these are coincident with near-IR cluster 
members listed in \citet{HC00}.  
An additional 21 objects were detected above the 3$\sigma$ 
level toward positions of near-IR sources.  1 mm continuum images of 
sources detected in our mosaic are displayed in Figure \ref{fig:detections}. 

While the BN object is detected, we defer discussion of this
high-mass, embedded object \citep[e.g.,][]{GBW98,PLAMBECK+95} to a 
later paper that examines the BN/KL region in detail\footnote{Although 
Source I is detected as a 
strong, individual object in our CARMA B-array data, we do not detect it in
our combined SMA+CARMA mosaic because of confusion with the hot core.  Source
I will also be discussed in the later paper.}.  
In the remaining discussion, we focus our attention on the sources detected
at both infrared and millimeter wavelengths.  While most of these  
objects do not have
known stellar masses, they are likely to be low-mass stars based on
the stellar mass distribution computed statistically for the ONC as a whole
\citep{HC00}.


\subsection{Angular Sizes of Detected Objects \label{sec:sizes}}
For each source detected in our 1 mm mosaic, 
we fitted a 2-D elliptical Gaussian
to the emission.  The synthesized, clean, beam for the combined mosaic is a
2-D Gaussian with FWHM of  $0\rlap{.}''69 \times 0\rlap{.}''60$ at a position
angle (north of west) of $72^{\circ}$.  At the assumed 400 pc distance to the
ONC, the core of the synthesized beam
has dimensions of 240 AU by 280 AU (again at a PA of $72^{\circ}$).
For simplicity, we approximate this as 240 AU in the East-West direction
and 280 AU in the North-South direction.  

For sources detected at a signal-to-noise ratio of $\sim 5$, the 
statistical uncertainty
in the fitted FWHM is $\sim 10\%$. Baseline errors or phase noise in our
mosaics can broaden the apparent source sizes, however.  We assume
that a source is resolved only if the major or minor axis 
of the fitted FWHM is 25\% larger than that of the synthesized beam.  
Objects for 
which the fitted Gaussian FWHM is smaller are considered to be unresolved.

Approximately 25\% (9/39) of detected sources are spatially resolved in 
our images (Table \ref{tab:results}).  
An additional nine objects (all proplyds) have been spatially
resolved with HST \citep{VA05}.  Thus, angular sizes are available for 
$\sim 50\%$ of our sources.  The inferred radii for
resolved sources range from $\sim 90$ to $\sim 220$ AU.
For unresolved objects we can say only that the emission is 
confined to radii smaller than\footnote{Larger sources, when
convolved with the synthesized beam, would produce measured sizes $>25\%$
broader than the beam.} $\sim 100$ AU.
For a sample of 134 proplyds with sizes measured with HST, 
the mean disk radius is 71 AU \citep{VA05}.  Since the mean disk diameter is 
$\sim 1/2$ the size of the linear resolution of our observations, it is not 
surprising that most of the sources detected in the 1 mm mosaic are unresolved.

\subsection{Distinguishing Dust and Free-Free Emission \label{sec:ff}}
Since we are interested in using our observations to constrain the mass of
circumstellar dust around our sources, we must account for potential 
contributions to the observed fluxes from sources other than dust emission.
Free-free emission arises in hot ionized gas, and in the ONC
such conditions may exist either in HII regions around high-mass stars
\citep[e.g.,][]{GMR87,PLAMBECK+95} or in the outer regions of disks or 
envelopes that are 
irradiated by the hot Trapezium stars \citep[e.g.,][]{OWH93,HO99}.
While some sources in the ONC have shown flares of cyclotron emission
\citep[e.g.,][]{BOWER+03,FURUYA+03}, 
we expect that such rare events will not contribute 
significantly to the 1.3 mm fluxes, and we do not consider them here.

Because the spectral shape of free-free radiation differs
from that of thermal dust emission, comparing
1 mm measurements with longer-wavelength data enables us to
distinguish these components.  
We use long-wavelength fluxes from the literature 
\citep{FELLI+93a,FELLI+93b,MLL95,ZAPATA+04,EC06,FMR07}.  
In addition, we use 880 $\mu$m fluxes measured by \citet{WAW05} for the few
objects where these are available.

For a freely expanding, fully ionized wind with constant 
$\dot{M}$, such as we expect for proplyds,
free-free emission will have the following spectrum\footnote{In a previous
paper \citep{EC06}, we assumed that free-free emission originated from
static HII regions rather than from winds, as in the present work.
This choice affects only the long-wavelength behavior of the free-free
spectrum, and is relatively unimportant to our analysis.}: 
\begin{equation}
F_{\rm \nu, ff} = \cases{F_{\rm \nu, turn} (\nu / \nu_{\rm turn})^{-0.1} &
if $\nu \ge \nu_{\rm turn}$ \cr
F_{\rm \nu , turn} \left({\nu} / {\nu_{\rm turn}}\right)^{0.6} &
if $\nu \le \nu_{\rm turn}$}.
\label{eq:ffwind}
\end{equation}
Here, $\nu_{\rm turn}$ is the frequency above which the wind is
optically thin at all radii.  
We include a derivation of this result
in the appendix, and an alternative derivation can be found in \citet{WB75}.

For $\nu < \nu_{\rm turn}$, the inner
parts of the wind are optically thick to free-free radiation.
If we adopt a simple model for proplyd winds where 
$\dot{M}=10^{-7}$ M$_{\odot}$ yr$^{-1}$ with spherical wind velocity of 20 km 
s$^{-1}$ \citep[e.g.,][]{HO99}, and $T_{\rm e} = 10^4$ K, we
can estimate the size of the optically thick region.  Using Equation
\ref{eq:xff}, we obtain $x_{\tau\approx 1} \sim 1$ AU at $\lambda$1 mm and 
$\sim 30$ AU at $\lambda$10 cm.  Even at 10 cm, this is smaller 
than the likely wind-launching regions
for proplyds, and free-free emission from most proplyds is likely to be fully 
optically thin.  For the highest measured mass loss rates of
$\sim 10^{-6}$ M$_{\odot}$ yr$^{-1}$ \citep[e.g.,][]{HO99}, the optically thick
regions of proplyd winds are $\sim 130$ AU for $\lambda \sim 10$ cm.
We therefore do not expect to see a spectral turnover  
(Equation \ref{eq:ffwind}) for wavelengths $\la 5$ cm.


Emission from cool dust is added to this free-free emission to obtain
a model of the observed flux.  We assume that 
\begin{equation}
F_{\rm \nu, dust} = F_{\rm 230 GHz, dust} (\nu / 230 {\rm \: GHz})^{(2+\beta)}
= F_{\rm 230 GHz, dust} (\nu / 230 {\rm \: GHz})^{3},
\label{eq:dust}
\end{equation}
for $\beta=1$ \citep[e.g.,][]{BECKWITH+90}.  Other values of $\beta$ can
not be ruled out based on our data for most objects, and $\beta=0$,
corresponding to emission from optically thick or large-grained dust,
is typically compatible with the data.

We estimate the relative contributions of dust and free-free emission
by fitting this model,  $F_{\nu} = F_{\rm \nu, ff}+F_{\rm \nu, dust}$, to our 
measured 1.3 mm fluxes and to 880 $\mu$m,  
3 mm, 3.6 mm, 1.3 cm, 2 cm, 6 cm, and 20 cm fluxes 
from the literature.  For comparison, we also fit a dust-only model,
described by Equation \ref{eq:dust}, with $\beta=1$ and $\beta=0$.
For objects detected at centimeter wavelengths, we fit the 
dust+free-free model to the $\ge 4$ flux measurements
for each source, and thus we are able to determine the three free parameters
of the model, $\nu_{\rm turn}$, $F_{\rm \nu, turn}$, and $F_{\rm \nu, dust}$.
For sources with $\le 3$ flux measurements (i.e., those undetected
in centimeter wavelength surveys), 
we fit only the dust emission model to the data.

Given the noise level of  previous centimeter observations covering
the entire region of our 1 mm mosaic 
\citep[$\la 0.3$ mJy;][]{FELLI+93a,FELLI+93b} and the measured
1 mm fluxes for detected objects ($\ga 10$ mJy), sources undetected at 
cm wavelengths are probably dominated by dust emission.  
For a source with 10 mJy flux at 1 mm, a non-detection at 10 cm
implies that $\la 0.2$ mJy, or $\la 2\%$ of the measured 1 mm flux is due
to free-free emission.  For simplicity, we attribute 100\% of the 1 mm fluxes 
to dust emission for these objects. 

\begin{figure}
\plotone{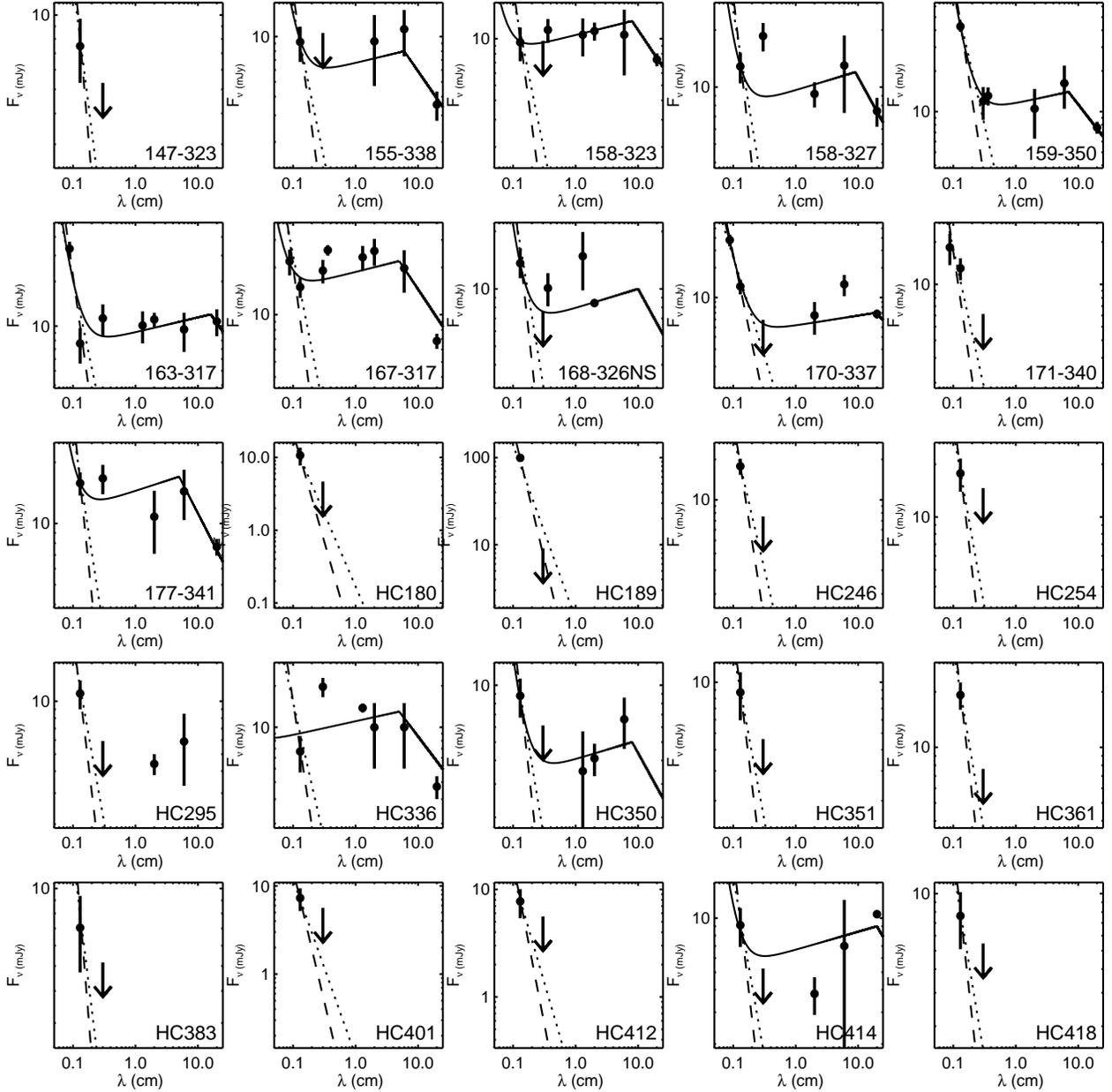}
\caption{Millimeter and radio fluxes for our sample (points), along with 
best-fit models including free-free and thermal dust emission.
Models including free-free and dust emission are indicated by solid lines, and
dotted and dashed lines show dust-only models with $\beta=0$ and $\beta=1$,
respectively.  The free-free flux density
is proportional to $\nu^{-0.1}$ for gas that is optically thin for all radii,
and to $\nu^{0.6}$ for a partially optically thick gaseous wind; the 
emission is
thus parameterized by the flux at a single wavelength and a turnover
frequency.  The emission from cool dust is proportional to $\nu^{2+\beta}$.
\label{fig:seds}}
\end{figure}

Fluxes, from sub-millimeter to radio wavelengths, and models are plotted in 
Figure \ref{fig:seds}, and the
fluxes due to thermal dust emission 
are listed in Table \ref{tab:results}.  Uncertainties for these 
dust fluxes are given by 
the 1$\sigma$ uncertainties of the model fits.
The majority of detected sources appear to be dominated by dust emission.  
However, for the subset of the sample seen in optical emission
or absorption with HST (the proplyds at the top of Table 
\ref{tab:detections}), the 1 mm fluxes are dominated by 
free-free emission.  This probably reflects the relative proximity of
proplyds to the luminous Trapezium stars.

\subsection{Estimating Circumstellar Dust Masses \label{sec:massest}}
The mass of circumstellar dust is related to the component of the 1 mm
continuum flux due to dust emission.   Assuming the dust is 
optically thin, and following \citet{HILDEBRAND83},
\begin{equation}
M_{\rm dust} = \frac{S_{\rm \nu,dust} d^2} 
{\kappa_{\rm \nu,dust} B_{\nu}(T_{\rm dust})}.
\label{eq:dustmass}
\end{equation}
Here, $\nu$ is the observed frequency,
$S_{\rm \nu,dust}$ is the observed flux due to cool dust, $d$ is the distance 
to the source,
$\kappa_{\rm \nu, dust} = \kappa_0 (\nu / \nu_0)^{\beta}$ is the 
dust mass opacity,
$T_{\rm dust}$ is the dust temperature, and $B_{\nu}$ is the Planck function. 
We assume $d \approx 400$ pc,
$\kappa_0=0.0002$ cm$^{2}$ g$^{-1}$ at 1.3 
mm, $\beta=1.0$ \citep{HILDEBRAND83,BECKWITH+90}, and $T_{\rm dust} = 20$ K
(based on the average dust temperature inferred for Taurus; Andrews \&
Williams 2005; see also the discussion in Carpenter 2002; 
Williams et al. 2005).
The dust mass can be converted into a total circumstellar mass by assuming
the canonical gas-to-dust mass ratio: 
$M_{\rm circumstellar} = M_{\rm dust} \times 100$.
Column 3 of Table \ref{tab:results} lists the estimated circumstellar
masses for detected objects.

Uncertainties in the assumed values of these parameters (notably $\kappa$) 
lead to uncertainties in the derived masses (in an absolute sense)
of at least a factor of $3$ \citep[e.g.,][]{POLLACK+94}, which are not
included in the uncertainties listed in the table.  Values of 
$\kappa_{\rm \nu,dust}$ and $T_{\rm dust}$ may also vary across our
sample.  Since the cluster population in Orion is roughly co-eval
\citep[e.g.,][]{HILLENBRAND97}, such effects should be minimal.  However,
there is some spread in stellar masses, which may lead to some range
in these parameters.  For example, since some of the objects in Table 
\ref{tab:detections} 
may be massive stars, the millimeter flux may contain contributions 
from dust hotter than the assumed 20 K;
if $T_{\rm dust}=30$ K, then the computed dust masses would be 
lower by a factor of 1.6.
For the predominantly low-mass cluster population \citep{HC00}, from which
the sources listed in Table \ref{tab:detections} are drawn,
the assumed values for $\kappa_{\rm \nu, dust}$ and $T_{\rm dust}$
presumably do not vary much, and the masses predicted by Equation 
\ref{eq:dustmass} are probably valid in a relative sense to within a
factor of two.  

\subsection{Constraints on Dust Optical Depth \label{sec:tau}}

We perform a simple test to determine whether the optical depth ($\tau$) 
might become 
comparable to or larger than unity in the systems under study.  
Using disk sizes measured from
either our data or HST data \citep{VA05}, or limits on disk sizes from
our observations, we compute the emission expected from optically thick
dust with a temperature of 20 K:
\begin{equation}
F_{\nu, \tau \ga 1} = B_{\nu} (T_{\rm dust}) 
\pi \left(\frac{R_{\rm disk}}{d}\right)^2 
\cos i \approx 112 {\rm \: mJy} \left(\frac{R_{\rm disk}}{\rm 100 \: AU}
\right)^2.
\end{equation}
Here $R_{\rm disk}$ is the disk radius, $i$ is the inclination and 
$d$ is the distance.  For simplicity, we take $i=0$, which leads to an
upper limit on the flux for an optically thick disk of radius $R_{\rm disk}$.
Fluxes expected for optically thick dust for our sample are
listed in Table \ref{tab:results}.

For most objects, the fluxes (or upper limits) expected for optically thick 
dust are substantially higher than our measured fluxes 
(Table \ref{tab:results}).  These disks
must be either highly inclined
or optically thin.  Since it is unlikely that all of the disks detected
in our observations are edge-on (especially since these would be the dimmest
portion of a sample of randomly inclined, optically thick disks), 
we take this as evidence for optically thin material.
Although many of the disk sizes, and hence the expected optically thick 
fluxes, are upper limits, the mean radius for proplyds in Orion of 
$\sim 70$ AU \citep[e.g.,][]{VA05} would produce an optically thick flux of 
$\sim 50$ mJy, still higher than the majority of our measured fluxes.

We therefore believe that most disks detected in our observation 
are composed largely of optically thin dust, and the circumstellar dust masses 
computed in \S \ref{sec:massest} are reasonable for these sources.
There are a few among the sources detected only at $\ga $ mm 
wavelengths (MM8, MM13, and MM21) for which the measured fluxes are 
comparable to or larger than the expected optically thick fluxes.  The
dust in these objects is either optically thick or hotter than 20 K,
as may occur around higher-mass (proto-)stars.  

\subsection{Stacking Analysis \label{sec:stacking}}
With the large number of young stars contained within our
mosaic, we can enhance the effective sensitivity by considering the
ensemble of $\sim 225$ sources not detected individually. 
For each known cluster member within the mosaic that is not detected
above the 3$\sigma$ noise level, we make a $10'' \times 10''$ sub-image
centered on the stellar position.  We weight the sub-image by the
local RMS (determined as described in \S \ref{sec:thresh}), sum all
of the weighted images, and divide by the sum of the weights to
normalize.  
We exclude any cluster members known to have radio-wavelength
emission \citep{FELLI+93a,FELLI+93b}.

\begin{figure}
\plotone{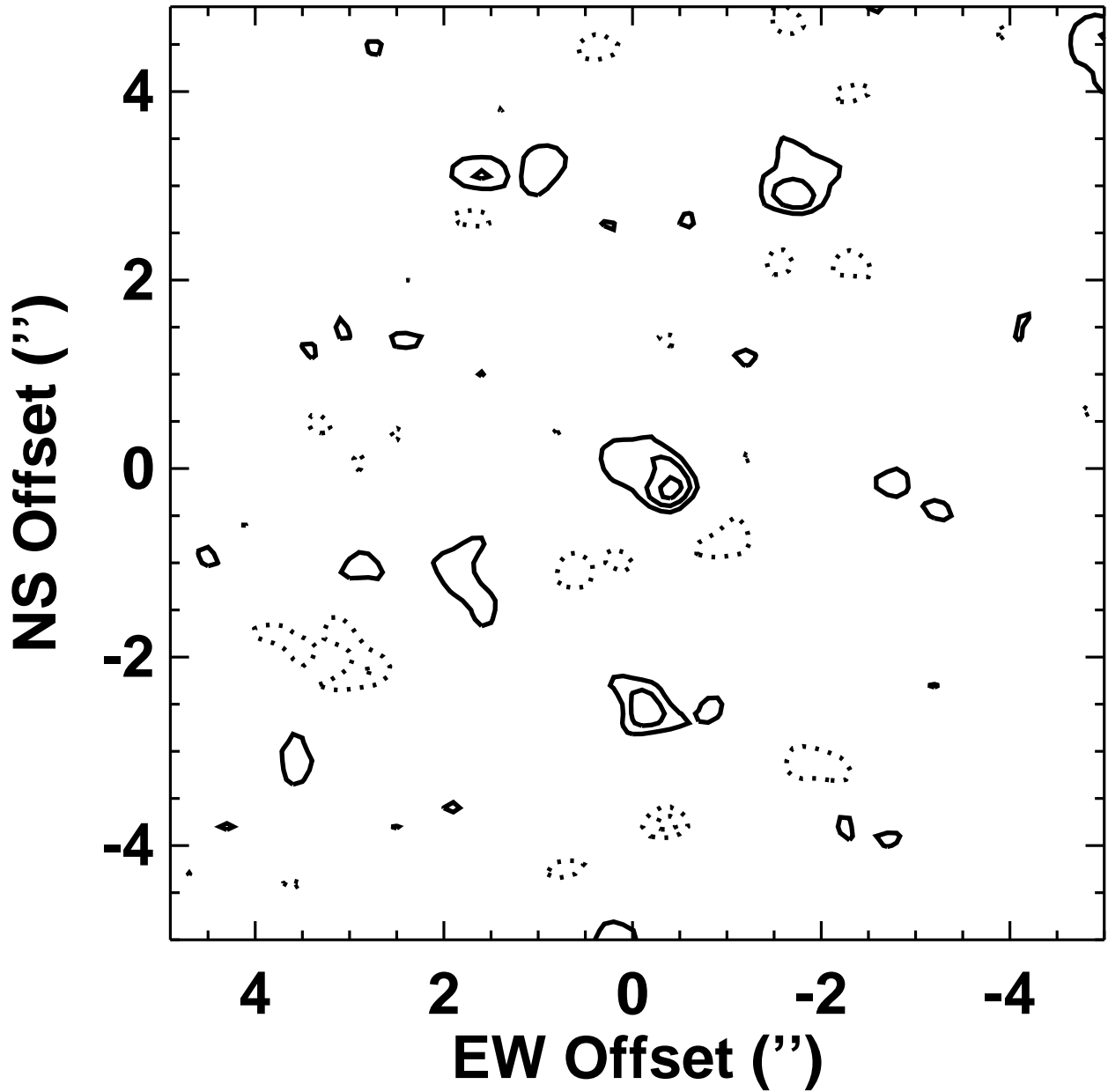}
\caption{
Average image, obtained by stacking the 1 mm continuum emission observed
toward each of 226 low-mass near-IR sources not detected 
individually above the 3$\sigma$ level.  Contour levels begin at 
$\pm 2 \sigma$ and the contour interval is $1 \sigma$ (negative contours
are shown as dotted lines).  Emission is detected 
for the ensemble at a significance of $\ga 4\sigma$, and exhibits
a compact (and beam-like) morphology approximately centered on the origin.
The degree to which the emission is smeared out is consistent with
the $\sim 0\rlap{.}''4$ positional errors in the near-IR source positions. 
\label{fig:avg}}
\end{figure}

The weighted image is shown in Figure \ref{fig:avg}.  The average flux 
for the ensemble of non-detected sources is 
$0.9 \pm 0.2$ mJy, with a significance of $> 4\sigma$.
The peak flux is centered on the mean position
of the near-IR sources (within the positional uncertainties of 
$\sim 0\rlap{.}''4$), and resembles the synthesized beam core, indicating 
that the average source is compact.  

Since the positional uncertainties are comparable to the half-width 
half-maximum of the synthesized beam, the average flux seen in Figure 
\ref{fig:avg} may be slightly reduced because different sources in the
ensemble do not lie exactly atop one another. Assuming
the positional uncertainties are random and Gaussian-distributed, one
would expect a reduction in  the measured average flux of $\sim 35\%$.
Correcting for the potential flux-smearing, one would obtain an average
flux for the ensemble of $1.2 \pm 0.2$ mJy.  We verify this by integrating
the central region of the average image over a region with four times
the area of the synthesized beam; as expected, we find an integrated
flux of $\sim 1.2$ mJy.

Low-level free-free emission may contaminate the average image, and hence
bias the average flux inferred for the ensemble.
The 1$\sigma$ sensitivity in cm-wavelength surveys is $\sim 0.3$ mJy
\citep[e.g.,][]{FELLI+93b}.
We argued in \S \ref{sec:ff} that gaseous winds in the ONC are likely to
be optically thin to free-free emission, and hence that 
$F_{\rm \nu, ff} \propto \nu^{-0.1}$.   Thus we would expect free-free emission
to be no stronger than $\sim 0.2$ mJy at 1 mm wavelengths for sources 
undetected in cm-wavelength surveys.  This is comparable to the 1$\sigma$
sensitivity in our 1.3 mm average image.  Since the average image is detected
above the 4$\sigma$ level, $>75\%$ of the average flux comes from dust.

\section{Discussion \label{sec:disc_o2}}

\subsection{Nature of Detected Sources \label{sec:geometry}}
\begin{figure}
\plotone{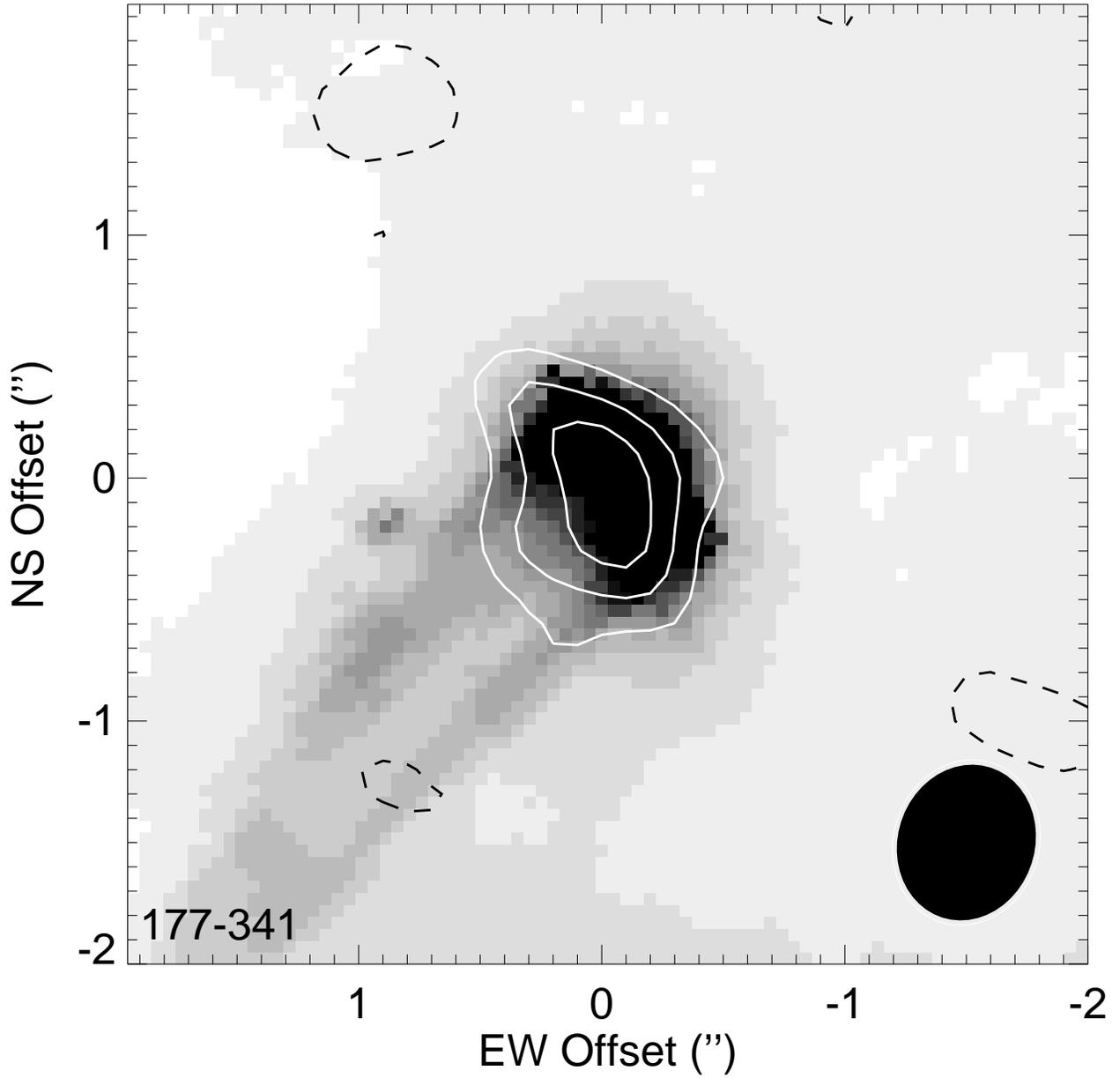}
\caption{Contour image of the 1.3 mm wavelength
continuum emission toward the proplyd 
177-341, overlaid on an HST map \citep{BOM00}
of the same position.   Positive contours
are white, solid curves and negative contours are dotted black curves.
Contours are $\pm 2,4$ and 6$\sigma$.  The FWHM of the synthesized beam is 
shown as a filled back symbol.  
\label{fig:proplyds}}
\end{figure}

The sub-arcsecond resolution of the CARMA observations is enough to 
marginally resolve some of the detected sources in the ONC, and in
principle we could observe flattened, disk-like geometries.
For example, Figure \ref{fig:proplyds} shows that 177-341 has a 
disk-like morphology aligned with the silhouette disk seen by HST.
While only a few sources can be well-resolved with our observations, 
HST observations show that many of the observed proplyds appear disk-like
\citep{MO96,BALLY+98a}, some even exhibiting silhouette disks \citep{BOM00}.
For the proplyds and well-resolved mm sources, the 1 mm  
emission evidently arises from disk-like distributions.

Mid-IR emission is also observed toward many of
the sources detected at 1 mm.  82\% of sources 
(all except the ``MM'' sources and HC 495) are also 
seen at 3.6 $\mu$m \citep{LADA+04}, and 48\% are seen at 11.7 $\mu$m
\citep{SMITH+05}.  While 3.6 $\mu$m emission may trace stellar
photospheres and/or infrared excess, the 11.7 $\mu$m emission
provides direct evidence for circumstellar material at least out to 
radii of a few AU.  Thus, many detected sources (the majority, if the ``MM'' 
objects are excluded) have evidence for inner circumstellar disks.

More generally, where 1 mm emission is detected
toward known near-IR cluster members, the fact that the near-IR light
is visible despite the high extinctions ($A_{\rm V} \ga 300$) that one 
would derive based on the amount of material needed to produce the 1 mm 
emission (for spherically distributed 
material) implies that the dust lies in flattened, disk-like 
distributions \citep[see also, e.g.,][]{BECKWITH+90,EC03}.  

It is interesting to speculate as to the nature of sources detected at
$HKL$ bands and at mm wavelengths, but not at 11.7 $\mu$m. 13/32 (40\%) of
mm and near-IR detected sources fall into this category.  
It is possible that some of these are transitional disks.
The $HKL$ emission may trace the stellar photosphere of a late-type star
while the mm emission traces a remnant outer disk, 
but large inner clearings may lead to a lack of mid-IR excess. 
Better coverage of the wavelength range between 10 $\mu$m and
1 mm is needed to test this hypothesis.


For sources without near-IR detections, the ``MM'' sources in 
Table \ref{tab:detections}, the arguments presented above do not apply.
Although the emission appears to trace circumstellar dust, the fact that 
no near-IR counterparts are observed suggests high columns of obscuring
material.   The MM sources all lack 
mid-IR counterparts as well.  These objects appear to be so embedded that they 
are still highly obscured even at 11.7 $\mu$m.   
All of the MM sources
reside in either the Orion BN/KL or OMC1-S region, both of which 
are known to contain young, embedded sources, HII regions, and outflows
\citep[e.g.,][]{ZWM90,BACHILLER96,ZAPATA+04}.  

All of the sources in OMC1-S (LMLA162, MM8, MM13, MM21, and MM22)
have been detected at 1.3 mm wavelength in previous observations 
\citep{ZAPATA+05,ZAPATA+07}.  Measured fluxes are similar to 
those listed in Table \ref{tab:detections}, but somewhat lower in most
cases, presumably because the poorer $uv$ coverage did not allow large negative
sidelobe contributions from extended emission to be fully removed.
All of these objects appear to drive molecular outflows
traced by CO or SiO emission \citep{ZAPATA+05,ZAPATA+06}.

We classify the MM sources as candidate Class 0 or Class I 
protostars.  As discussed above, it appears that the 1.3 mm emission from at 
least some of these sources may 
trace dust hotter than 20 K.  Such warm dust is expected in the 
circumstellar environments of massive protostars, suggesting that some of
the MM sources trace high-mass protostars. 

Several sources detected in previous surveys were not detected here.  
HC178, HC192, HC282, MM3,  MM4, MM10, MM15, MM16, MM19, and MM20
should have been detected if their 3 mm fluxes \citep{EC06} traced dust
emission; however, they would not have been detected if the objects exhibited 
flat spectra (e.g., from free-free emission).  MM7, MM17, and MM18 
should have been detected even if they showed flat spectra.
We detect a 1.3 mm
continuum source (MM21) near to HC 178, but find it to be offset by more than
the relative positional uncertainties, suggesting that the previous
association of HC 178 with a 3 mm source was mistaken.  
The other 3 mm objects trace either non-dust, potentially time-variable
emission, or are spurious, caused by confusion with the BN/KL and OMC1-S
regions in which they reside.  Because our 1 mm observations have far 
superior $uv$ coverage than previous observations, they are less prone
to such spurious detections.  One source detected by \citet{WAW05} at 
880 $\mu$m (171-334) is not detected at 1.3 mm; if the emission comes
from small-grained dust, then the expected 1.3 mm flux is comparable
to our 3$\sigma$ noise level, and hence a non-detection is unsurprising.

\subsection{Frequency of Massive Disks \label{sec:disc_mass}}
We detected 39 sources in our 1 mm mosaics (excluding the BN object).  
32 of these correspond to (presumed) low-mass near-IR cluster members, and 
6 (the ``MM'' sources) are detected only at $\ga 1$ mm wavelengths.  
The remaining detection, LMLA162, while not listed as a near-IR source in 
\citet{HC00}, is seen at 3.6 $\mu$m \citep{LADA+04}; 
examination of an archival 2MASS
image shows a weak 2 $\mu$m source at this position as well.
Since the mm-only detections are probably 
embedded, possibly spherical, protostellar objects (\S \ref{sec:geometry}),
we exclude these from our discussion of disk statistics.  Of the remaining
33 detections, 100\% of the 1 mm emission can be attributed to hot gas 
(free-free) for 6 sources.  Thus, we are left with 27 sources whose 1 mm 
emission (probably) traces dust in protoplanetary disks.

Since the noise varies across our images, these 27 sources are
all detected above slightly different thresholds.  To examine the frequency
of disks more massive than some value, we make sensitivity cuts at various
levels, examining only the statistics of sources detected above chosen
noise levels.  

We consider first the 115 cluster members surveyed to a 1$\sigma$ noise 
level of 2.7 mJy or less.  Sources detected above 3$\sigma$ have
a circumstellar (dust+gas) mass of $\ge 0.01$ M$_{\odot}$.
Nine sources (8\%) show dust emission of $\ge 8.1$ mJy 
(i.e., 3$\sigma$ detections at this noise level).  
If we use a higher noise cutoff of 5 mJy, then we 
find that 7 out of 193 stars  exhibit dust emission 
above the 3$\sigma$ level of 15 mJy.  So, $\sim 4\%$ 
of stars have disks more massive than 0.02
M$_{\odot}$.  If we extend the sensitivity cutoff further, to 10 mJy,
then 3/254, or $\sim 1\%$ of stars are seen with disks more massive than 0.04
M$_{\odot}$.  
All of the 3$\sigma$ mass levels considered here fall within the 
range of estimates for the MMSN 
\citep[$\sim 0.01$--0.1 M$_{\odot}$;][]{WEID+77}.  

The percentage of high-mass disks derived here can be compared to that
determined by \citet{EC06}.  The observations presented here are
substantially more sensitive than previous observations, and  we
probe the frequency of disks down to lower mass levels; we can therefore
only compare statistics for the most massive disks in our sample.  \citet{EC06}
found that $\le 2\%$ of cluster members in the ONC have disks more massive than
0.1 M$_{\odot}$.  They assumed a distance of 480 pc; their mass limit is 
actually only 0.07 M$_{\odot}$ for the distance of 400 pc assumed here.
Here we find that $\la 1\%$ of stars surveyed are surrounded by 
such massive disks, consistent with the estimate from \citet{EC06}.

We emphasize that that results presented above (and in the following
sections) depend on the conversion of 1.3 mm flux into mass.  As
discussed in \S \ref{sec:massest}, there may be some spread in the
dust properties of our sample that could lead to variations in the
derived circumstellar masses.  For the roughly co-eval, predominantly low-mass 
cluster population in the ONC, we argued that this is a relatively small
uncertainty.  

\subsection{A Typical Disk in the ONC \label{sec:avg}}
We computed the average
flux for the ensemble of non-detected sources in \S \ref{sec:stacking}.
The average flux indicates that a ``typical'' non-detected source in the ONC
likely possesses a disk with a mass of $0.0015 \pm 0.0003$ M$_{\odot}$.
If we include detected objects (whose dust fluxes are listed in
Table \ref{tab:results}) in the ensemble, we find that the
average disk mass for near-IR cluster members in the region is 
$\sim 0.0027 \pm 0.0002$ M$_{\odot}$.

This is comparable to the average mass determined for 23
proplyds in the ONC at 880 $\mu$m \citep{WAW05}, but substantially 
lower than the average mass determined for $>300$ stars at 3 mm wavelengths, 
$0.005 \pm 0.001$ M$_{\odot}$ \citep{EC06}.  The discrepancy can be 
explained  in large part by contamination from free-free emission.
Because the inferred dust mass is proportional to $\lambda^3 S_{\nu}$
(Equation \ref{eq:dustmass}), this contamination has a much greater affect on 
the masses inferred from the 3 mm data than on our estimates based on 
1 mm data.  The dust mass attributed to 
free-free emission is $(2.3)^3 \approx 12$ times larger at 3 mm than at 
1.3 mm.  If free-free emission is present at the $\la 0.2$ mJy level 
(\S \ref{sec:ff}) it would add $\la 0.003$ M$_{\odot}$ to the average mass 
inferred at 3 mm. Furthermore, if we recompute the mean disk mass from 
\citet{EC06} using a distance of 400 pc, the average mass is decreased
by 30\%.  With the distance correction and the subtraction of
potential free-free contamination, the recomputed average mass
from \citet{EC06} is $\ga 0.001$ M$_{\odot}$, 
in agreement with the estimated average mass inferred from our 1 mm 
observations.

\subsection{Comparison of Disk and Exoplanet Frequencies \label{sec:exop}}
Less than $10\%$ of stars in 
the ONC possess disks comparable to the MMSN 
(\S \ref{sec:disc_mass}).  
Moreover, the
average mass measured for the ensemble of (individually) non-detected sources 
is ten times smaller than even the low end of estimates for the MMSN
(\S \ref{sec:avg}),
indicating that the majority of stars do not possess enough mass to form 
Jupiter-mass planets.

These statistics can be compared with the frequency of Jupiter-mass planets
found around nearby main-sequence stars.  6\% of stars surveyed have a 
Jupiter-mass (or larger)
planet within 5 AU, while an extrapolation based on current
results suggests up to 10\% of stars could have a Jupiter-mass planet within
20 AU \citep{MARCY+05}.  The frequency of massive planets is comparable
to the frequency of disks in the ONC with (low-end) 
minimum minimum mass solar nebulae.
It appears that the MMSN, applied to disk mass measurements like those
presented here, is a reasonable criteria
for forming massive, Jupiter-like planets in typical star forming regions
like Orion.

\subsection{Disk Evolution \label{sec:disc_evol}}
The frequency of massive disks in the ONC (aged $\sim 1$ Myr)
can be compared with surveys of
rich clusters of different ages, NGC 2024 (aged $\sim 0.3$ Myr)
and IC 348 (aged $\sim 2$ Myr), 
to constrain the evolution of disks in
clustered star forming environments.  While this comparison has been made
previously using 3 mm observations \citep[e.g.,][]{EC06}, our 1 mm measurements
are less contaminated by free-free emission and yield 
different results (\S \ref{sec:avg}).  
Although the surveys of NGC 2024 and IC 348 were at 
3 mm, the lack of O stars in those regions should produce less ionized
gas, and hence less contamination by free-free emission than in the ONC.

\begin{figure}
\plotone{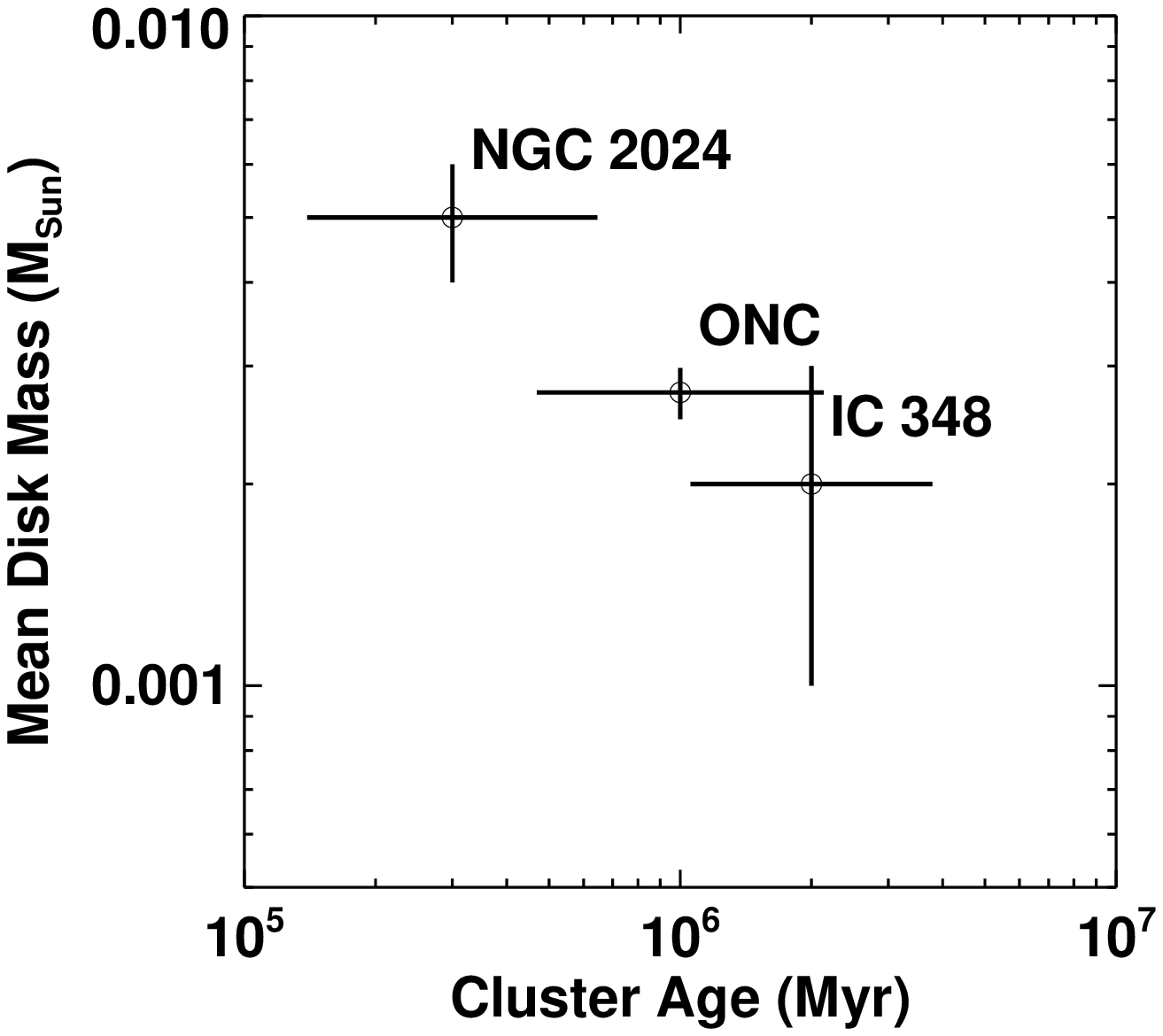}
\caption[Disk mass, as a function of cluster age]
{Average disk mass as a function of age for the NGC 2024, ONC, and
IC 348 clusters.  The disk masses are taken from this work, \citet{EC03},
and \citet{CARPENTER02}, and estimated cluster ages and uncertainties are from
\citet{MEYER96}, \citet{ALI96}, \citet{HILLENBRAND97}, \citet{LUHMAN+98},
and \citet{LUHMAN99}.  Average disk masses for NGC 2024 and IC 348 were
measured at 3 mm, where potential contributions from free-free
emission would be stronger than for the average mass measured here for
the ONC at 1 mm.  We argue in \S \ref{sec:disc_evol}, however, that free-free
contamination is unlikely in NGC 2024 and IC 348.
\label{fig:evol}}
\end{figure}

The average disk masses for ``typical'' low-mass stars in the three regions
is plotted as a function of cluster age in Figure \ref{fig:evol}.
For the ONC, we infer a mean disk mass of $0.0027 \pm 0.0002$
(\S \ref{sec:avg}).  
In NGC 2024, the mean disk mass is $0.005 \pm 0.001$ M$_{\odot}$
\citep{EC03}, compared to $0.002 \pm 0.001$ M$_{\odot}$ in IC 348 
\citep{CARPENTER02}.   If the differences
between NGC 2024, the ONC, and IC 348 are due to temporal evolution, these
observations suggest that massive disks/envelopes dissipate on timescales
$\la 1$ Myr, and that the average disk mass 
decreases by a factor of $1.9 \pm 0.4$ between $\sim 0.3$ and 
1 Myr.  



\subsection{Dependence of Disk Properties on Environment \label{sec:disc_env}}
It has been suggested that circumstellar disks in clustered environments
may be truncated due to close encounters with massive stars resulting
in either tidal stripping or photo-evaporation of outer disk material
\citep[e.g.,][]{SC01}.  Indeed, photoevaporative mass loss has been
observed from many proplyds, suggesting mass loss rates as high as
$10^{-7}$--$10^{-6}$ M$_{\odot}$ yr$^{-1}$ 
\citep[e.g.,][]{HO99}, which would severely 
deplete the masses of disks over the $\sim 1$ Myr lifetime of the cluster.
More detailed models have shown that the mass loss rate should be substantially
lower for disks with smaller outer radii, since disk material at smaller
radii is more tightly gravitationally bound \citep{CLARKE07}.  A
prediction of these models is that larger disks will also be the most
massive, since they have to withstand higher photoevaporative mass
loss rates.  

The proplyds detected in our observations are in the top
$\sim 1/3$ of the size distribution inferred by \citet{VA05}.  However,
the emission from most of these is dominated by free-free emission,
and even for objects where some component of the flux is due to dust,
inferred masses are $\la 0.01$ M$_{\odot}$.
Furthermore, there are many other proplyds whose diameters are in the
top 30\% that are not detected in our observations.  We also see no
obvious trend of increasing flux with increasing angular size in our data
(Table \ref{tab:results}).  Thus, we
find little evidence that the most extended disks are the most massive.

\begin{figure}
\plotone{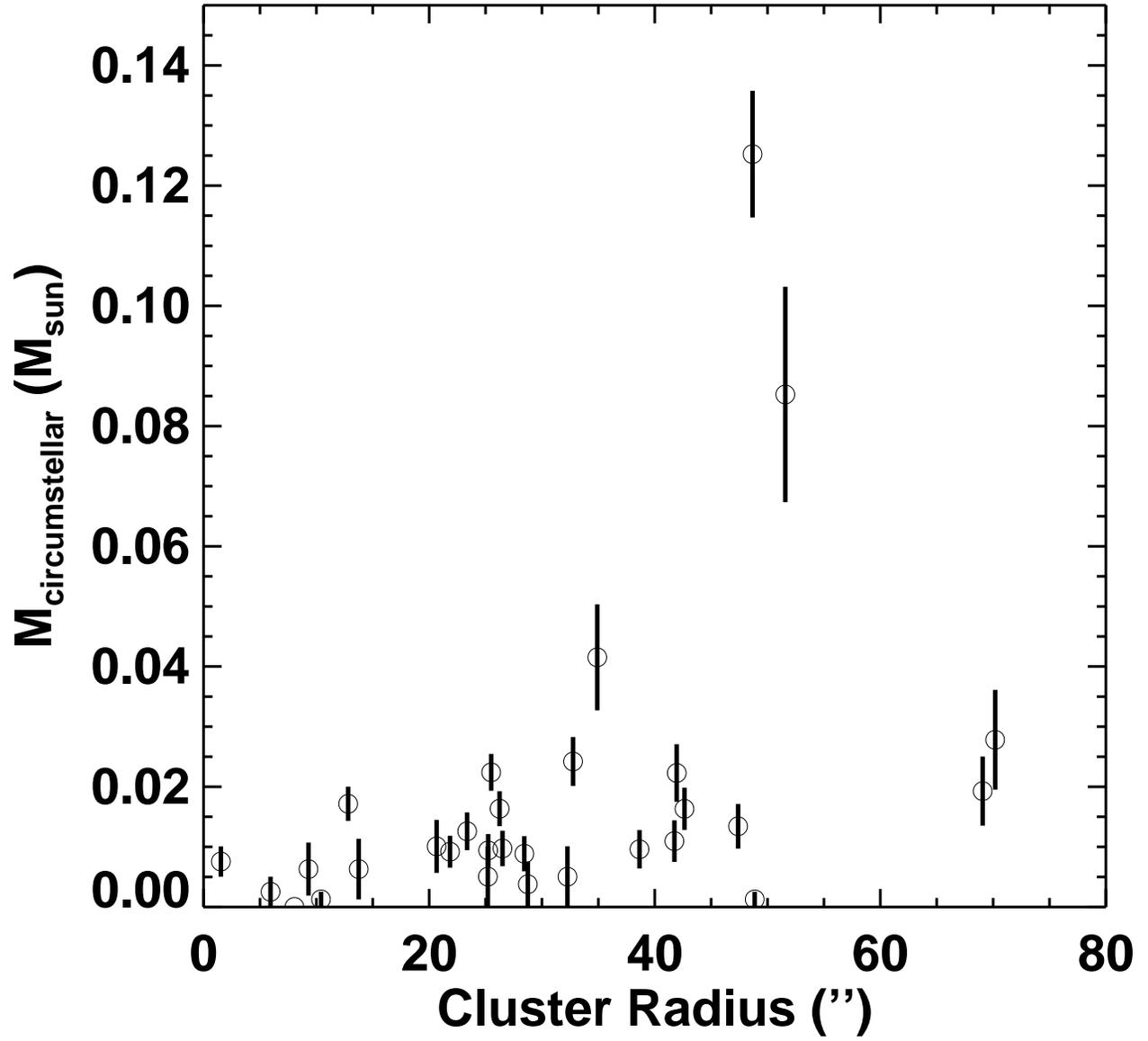}
\caption{Circumstellar mass as a function of radial distance from the
center of the Trapezium stars for detected objects.
\label{fig:radii}}
\end{figure}

Environmental effects on massive disks can also be investigated through
the dependence of disk properties on cluster radius.  
We consider the positions of the disks detected in our observations
(i.e., detected sources corresponding to known near-IR cluster member
positions) relative to the cluster center, which we define to lie
roughly in the middle of the four bright Trapezium
stars at $(\alpha,\delta)_{\rm J2000} = (5^{\rm h}35^{\rm m}16.34^{\rm s},
-5^{\circ}23'15\rlap{.}''6)$.   Figure \ref{fig:radii}
shows that more massive disks tend to be found further away from the
Trapezium stars.  If we consider only those 194 cluster members where we could 
have detected disks more massive than 0.02 M$_{\odot}$ above the 3$\sigma$ 
noise level, we find disks around 1/84 stars 
($\sim 1\%$ ) within $30''$ and 6/110 stars ($\sim 5\%$) at radii larger than 
$30''$.  Fisher's exact test indicates 86\% probability (1.5$\sigma$) 
that the small and large cluster radii sources have different 
frequencies of massive disks. 

Finally, comparison of the ONC with the lower stellar density Taurus region
provides another test of whether the massive O stars 
and high stellar density in the Trapezium
region lead to different disk properties than in more ``benign'' environments.
As discussed in \S \ref{sec:disc_mass}, we
detected disks more massive than $0.01$ M$_{\odot}$ 
around 9/115 ($\sim 8\%$) low-mass ONC cluster members. 
For comparison,  34/153 ($\sim 22\%$) of Taurus stars possess such
massive disks \citep{AW05}.   Fisher's Exact Test
yields $> 99\%$ probability ($3\sigma$) that the frequencies 
of 0.01 M$_{\odot}$ disks in Taurus and the ONC are different.
For a slightly higher mass cutoff of 
0.04 M$_{\odot}$, such massive disks are found around $< 1\%$ of
stars in the ONC compared to $\sim 5\%$ for Taurus.  These percentages
indicate $>99\%$ probability that the underlying distribution of 0.04
M$_{\odot}$ disks in Taurus and the ONC differ.
The fraction of approximately MMSN-massed disks in Orion is substantially 
smaller than in Taurus, arguing that the rich cluster environment may play a 
role in limiting the number of massive disks.  

This conclusion differs
from that of \citet{EC06}, where the statistics of disks more massive than
0.1 M$_{\odot}$ were found to be statistically indistinguishable in Taurus
and the ONC.  Using the revised distance of 400 pc changes the conclusion
from \citet{EC06}, because if statistics of (distance-corrected)
0.07 M$_{\odot}$ disks are
compared, they are found to be substantially more common in Taurus
(Fisher's Exact Test indicates only $\sim 1.3 \%$ probability that the two
distributions are the same).  Furthermore, the 1 mm observations
presented here are sensitive to much more of the disk mass distribution,
allowing a more robust comparison between Taurus and the ONC.

\subsection{Correlation of Circumstellar and Stellar Masses \label{sec:mstars}}
Several spectroscopic 
surveys have provided accurate masses for a subset of the stellar population
encompassed by our mosaics 
\citep{HILLENBRAND97,LUHMAN+00,SLESNICK+04}.  By examining these
surveys, after registration of the positions of detected sources to
the 2MASS grid, we find $\sim 130$ objects with spectroscopically-determined
masses within the unit gain contour of our mosaic.  While the stellar masses
of the remaining cluster members contained in our mosaic have
been estimated statistically by de-reddening stars so that they fall on
the expected isochrone for the ONC \citep{HC00} the masses of individual
stars determined in this way have large uncertainties and we do not use them 
here.

To examine how disk mass depends on stellar mass, we divide
the $\sim 130$ sources with spectroscopically-determined masses into 
three mass bins containing roughly equal numbers of stars.  
The first bin contains stars less massive than
0.3 M$_{\odot}$, the second bin includes stars with masses between 0.3 and
1.0 M$_{\odot}$, and the third bin contains stars with masses between
1.0 and 10.0 M$_{\odot}$.  We then make a further cut by excluding all
objects for which the noise in our 1 mm mosaic at the source position
is greater than some cutoff value.  As in \S \ref{sec:disc_mass}, we 
consider a noise cutoff of 2.7 mJy, which provides a corresponding
3$\sigma$ circumstellar mass threshold of 0.01 M$_{\odot}$.
Ideally, we would also bin this sample by cluster radius to control
for potential mass segregation in the inner regions of the ONC 
\citep[e.g.,][]{HILLENBRAND97}.  Unfortunately we lack a sufficiently large 
sample to do this here.

If we use the raw image fluxes (Table \ref{tab:detections}), we find 
more sources detected in the highest stellar mass bin.  For 65 stars with 
spectroscopically-determined masses, surveyed with a noise level of 2.7 mJy
or lower, we detect 1/21 (5\%) stars with 
$M_{\ast}<0.3$ M$_{\odot}$, 2/24 (8\%) stars with 
0.3 M$_{\odot} < M_{\ast}<1$ M$_{\odot}$, and 6/20 (30\%) stars with 
$M_{\ast}>1$ M$_{\odot}$.  However, the higher percentage of detected
sources in the highest stellar-mass bin is due entirely to contamination
by free-free emission: if we use dust fluxes from Table \ref{tab:results},
then 0/20 stars in the highest stellar mass bin are detected.  
We infer, therefore, that higher mass stars are more likely to
exhibit free-free emission.

\begin{figure}
\plotone{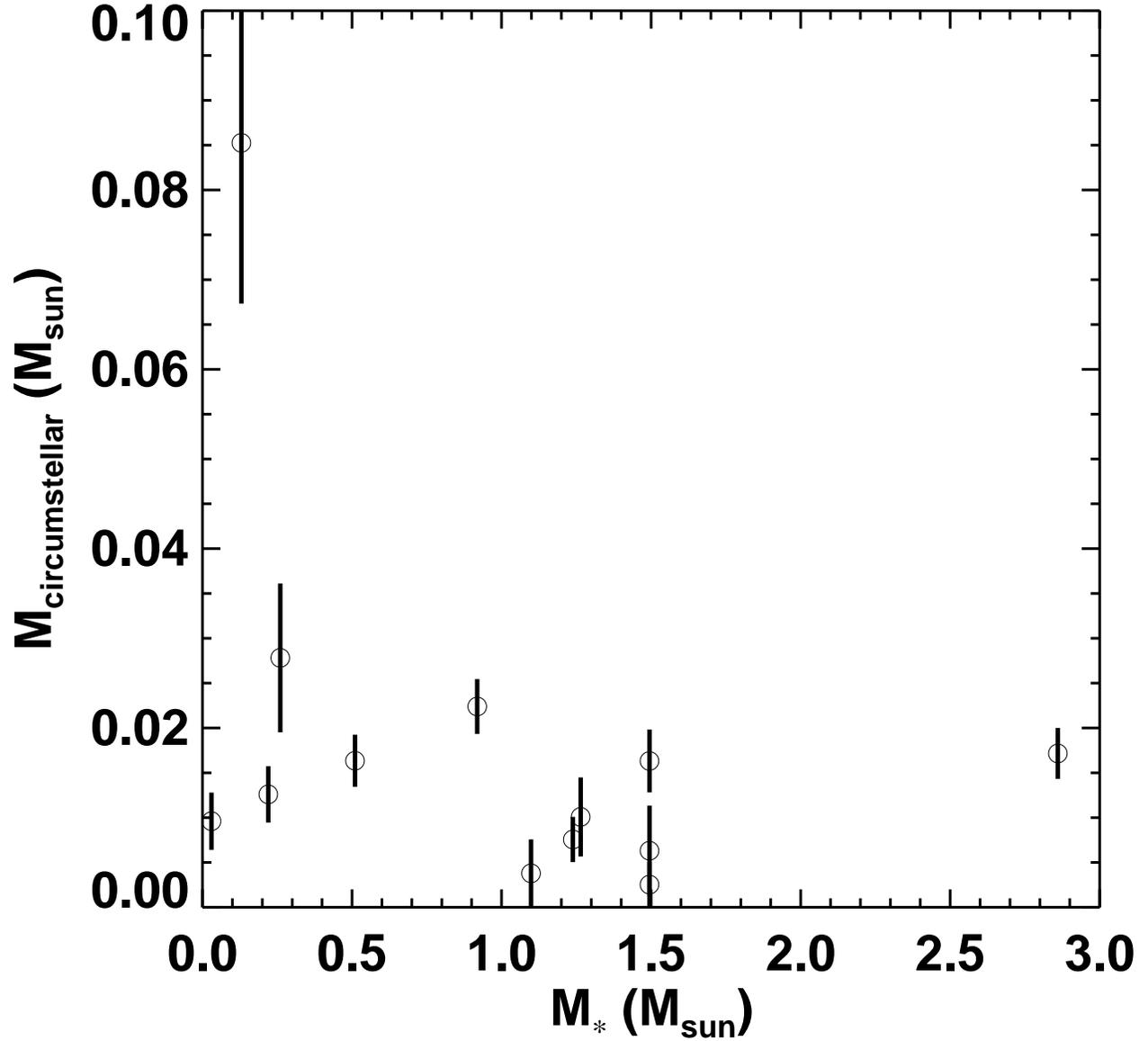}
\caption{Circumstellar mass as a function of stellar mass for the
subset of detected objects where spectroscopically determined stellar masses
are available \citep{HILLENBRAND97,LUHMAN+00,SLESNICK+04}.
\label{fig:mstars}}
\end{figure}

In contrast, it seems that more massive stars may be less likely to
possess massive circumstellar disks.  Using the dust fluxes from Table 
\ref{tab:results}, we find that out of the 65 stars discussed above, 
we detect {\it dust} emission above the 3$\sigma$ level toward
1/21 (5\%) stars with $M_{\ast}<0.3$ M$_{\odot}$, 2/24 (8\%) stars with 
0.3 M$_{\odot} < M_{\ast}<1$ M$_{\odot}$, and 0/20 (0\%) stars with 
$M_{\ast}>1$ M$_{\odot}$.  This suggests a lower frequency of disks
around stars more massive
than 1 M$_{\odot}$, but these small number statistics do not allow
a definite conclusion.  A plot of  inferred circumstellar disk masses versus 
stellar masses (where available) supports the hypothesis 
that the most massive disks are found
around the lowest mass stars.  This trend is not, however, statistically
significant for the small number of objects in Figure \ref{fig:mstars}.
A similar picture was seen in Taurus \citep{AW05}: no 
correlation between stellar mass and disk mass could be established, 
although the most massive disks were found around stars less massive than
1 M$_{\odot}$.  We note that \citet{NGM00} claimed to see
a correlation between disk and stellar masses around early-type stars;
however, the dispersion is large and the significance of the trend is marginal.


\section{Conclusions \label{sec:conc}}
We imaged a $2' \times 2'$ region of the Orion Nebula cluster in 1.3 mm
wavelength continuum emission with CARMA and the SMA.  Out of $>250$ known
near-IR cluster members, we detected 1.3 mm emission above the $3\sigma$ 
noise level toward 33.  In addition, we detected 1 mm emission above the 
5$\sigma$ noise level from six sources not associated with 
shorter-wavelength counterparts.
Several of these detected objects are spatially resolved with our observations,
indicating sizes of $\sim 250$--450 AU.  

Modeling of long-wavelength fluxes for our targets allowed separation of
dust and free-free emission components in the measured fluxes. 
We showed that for the majority of detected sources, the 1 mm emission appears 
to trace warm, optically thin, 
circumstellar dust.  However, for many of the proplyds, which are
located close to the Trapezium stars, the millimeter wavelength emission 
is dominated by thermal free-free emission from hot, ionized gas.

Dust masses inferred for detected sources range from 0.01 to 0.5 M$_{\odot}$.
For the $\sim 225$ known near-IR cluster members not detected in our 1 mm 
observations, images toward the positions of near-IR sources 
were stacked to constrain the mean flux, and circumstellar mass, 
of the ensemble.  The average flux is detected at the $> 4\sigma$ 
confidence level, and implies an average disk mass of $\sim 0.001$ 
M$_{\odot}$, approximately an order of magnitude smaller than the 
minimum mass solar nebula.  Even when detected sources are included,
the average mass is $<0.003$ M$_{\odot}$.
While the derived masses are uncertain by
a factor of 3 or so (mostly due to uncertainties in the dust opacity),
the range of possible average disk masses is still smaller than even the
low end of estimates for the MMSN.   A ``typical'' star in the ONC  does 
not appear to possess sufficient mass in small dust grains to form 
Jupiter-mass (or larger) planets.  Evidently, giant planet formation
is either advanced (having thus depleted the small dust grains in the disk) 
or impossible around most stars in the ONC.

We compared the average disk mass inferred for the ONC with similarly 
determined average masses in older and younger clusters.  We find evidence for 
evolution of the dust (most likely depletion or agglomeration) on
$\sim 1$ Myr timescales.  Between $\sim 0.3$ Myr and $\sim 1$ Myr, the
average disk mass decreases by a factor of $1.9 \pm 0.4$.

The percentage of stars in Orion surrounded by disks more massive than the
minimum mass solar nebula is $<10\%$.  This is significantly lower than in 
Taurus, indicating that environment has an impact on the disk mass 
distribution.  Our data suggest (with marginal statistical
significance) that the most massive disks may be located 
further from the Trapezium stars, supporting the hypothesis that 
photoevaporation may be truncating disks near to the cluster center.

Finally, our observations show no clear correlation between stellar mass and 
disk mass, but suggest that massive disks may be more likely
to be found around lower mass stars.  The percentage of detected disks
is lower for stars more massive than 1 M$_{\odot}$, and the most massive
disks detected are associated with the relatively low stellar mass stars
in the sample.
However, larger numbers of stellar and disk mass measurements in the ONC
are needed to build up better statistics and further
constrain the relationship between stellar and disk properties.

\acknowledgements
JAE acknowledges support from a
Miller Research Fellowship.  The authors thank M. Wright for his assistance
in simulating CARMA mosaics in order to choose the optimal mosaic
spacings for the observations.  This publication makes use of data products 
from the Two Micron All Sky Survey, which is a joint project of the 
University of Massachusetts and the Infrared Processing and Analysis Center 
(IPAC), funded by the National Aeronautics and Space Administration and the 
National Science Foundation. 2MASS science data and information services were 
provided by the Infrared Science Archive at IPAC.
Support for CARMA construction was derived from the states of Illinois, 
California, and Maryland, the Gordon and Betty Moore Foundation, the Kenneth 
T. and Eileen L. Norris Foundation, the Associates of the California Institute 
of Technology, and the National Science Foundation. Ongoing CARMA development 
and operations are supported by the National Science Foundation under a 
cooperative agreement and by the CARMA partner universities.


\appendix
\section{Free-free emission spectrum for an ionized wind \label{sec:ffapp}}

We compute the spectrum expected for free-free emission from ionized gas
whose density depends on stellocentric radius as 
\begin{equation}
n_{\rm gas} = n_0 (R/R_{\rm 0})^{-\alpha}.
\label{eq:ffdens}
\end{equation}
For an ionized wind with constant $\dot{M}$, such as we expect for proplyds, 
$\alpha=2$.  However, massive stars may ionize their circumstellar environments
directly, in which case the density distribution may differ from this
$R^{-2}$ power-law.  

For a power-law density profile (Equation \ref{eq:ffdens}), the optical depth
can be approximated as \citep[e.g.,][]{ALTENHOFF+60},
\begin{equation}
\tau_{\rm \nu,ff} \approx \int_{0}^{\infty}
\frac{0.16 n_{\rm 0}^2}{\nu^{2.1}T_{\rm e}^{1.35}} \frac{dz}
{(x^2+z^2)^{\alpha}}.
\end{equation}
Here, $z$ is the line of sight through the ionized gas in pc, $x$ is the
impact parameter in pc, $n_0$ is the normalization of the gas density
in cm$^{-3}$ (assumed to represent the electron and ion densities), 
$T_{\rm e}$ is the electron temperature,
and $\nu$ is the frequency in GHz.  This expression can be integrated
directly as long as $\alpha \ge 1.5$, with the result
\begin{equation}
\tau_{\rm \nu,ff} \approx \frac{0.08 n_{0}^2 R_0^4}{\nu^{2.1}T_{\rm e}^{1.35}} 
x^{-(2\alpha -1)}
\frac{\pi (-1)^{\alpha-1} \: \Gamma(1/2)}{\sin(\pi/2) (\alpha-1)! 
\: \Gamma(1/(2-\alpha+1))}.
\end{equation}
The last fraction has order unity, and we can thus approximate $\tau$ as
\begin{equation}
\tau_{\rm \nu,ff} \approx \frac{0.08 n_{0}^2 R_0^4}{\nu^{2.1}T_{\rm e}^{1.35}} 
x^{-(2\alpha -1)}.
\label{eq:fftau}
\end{equation}

We can now invert Equation \ref{eq:fftau} to determine the maximum impact
parameter for which $\tau \ga 1$:
\begin{equation}
x_{\tau\approx 1} \approx \left(\frac{0.08 n_0^2 R_0^4}{\nu^{2.1} 
T_{\rm e}^{1.35}}\right)^{\frac{1}{2\alpha-1}}.
\label{eq:xff}
\end{equation}
For impact parameters larger than $x_{\tau\approx 1}$, the gas is
optically thin, and at smaller impact parameters the gas is optically thick.
The total spectrum of free-free emission from the source can be approximated
by the blackbody flux times the solid angle of the optically thick region,
as long as the optically thick region is finite.  When the entire wind
becomes optically thin, the spectrum flattens.

We can thus parameterize the free-free emission from a wind as
\begin{equation}
F_{\rm \nu, ff} = \cases{F_{\rm \nu, turn} (\nu / \nu_{\rm turn})^{-0.1} &
if $\nu \ge \nu_{\rm turn}$ \cr
F_{\rm \nu , turn} \left({\nu} / {\nu_{\rm turn}}\right)^{\frac{4\alpha-6.2}
{2\alpha-1}} &
if $\nu \le \nu_{\rm turn}$}
\label{eq:ff}
\end{equation}
\citep[see][for a somewhat different derivation of this result]{WB75}.
For a spherical wind with constant $\dot{M}$, $\alpha=2$ and the 
optically thick
material emits with $F_{\nu} \propto \nu^{0.6}$.  Steeper exponents can result
if the ionized gas density drops off more steeply than $R^{-2}$, for example
as might occur in a centrally illuminated wind from a massive star like BN
\citep[e.g.,][]{PLAMBECK+95}.
Alternatively, non-spherical wind geometries can lead to shallower 
radial density profiles, and hence shallower spectral slopes of the free-free
emission \citep{WB75}.  For simplicity, since we generally have a limited
number of data points with which to constrain the free-free emission spectrum,
we will assume $\alpha=2$ in the present analysis.

\clearpage
\begin{deluxetable}{lccccccccccc}
\tabletypesize{\scriptsize}
\rotate
\tablewidth{0pt}
\tablecaption{Long-wavelength fluxes for sources detected in $\lambda$1.3 mm 
continuum \label{tab:detections}}
\tablehead{\colhead{ID} & \colhead{$\alpha$} & 
\colhead{$\delta$} & \colhead{$S_{\rm 880 \mu m}$ } & 
\colhead{$S_{\rm 1.3mm}$ } & \colhead{$S_{\rm 3mm}$ } & 
\colhead{$S_{\rm 3.6mm}$ } & \colhead{$S_{\rm 1.3 cm}$ } & 
\colhead{$S_{\rm 2cm}$ } & \colhead{$S_{\rm 6cm}$ } & 
\colhead{$S_{\rm  20cm}$ } \\
 & (J2000) & (J2000) & (mJy) & (mJy) & (mJy) & (mJy) & (mJy) & (mJy) & (mJy) 
& (mJy)}
\startdata
147-323 & 5 35 14.73 & -5 23 22.91 & $ $ & $  7.5 \pm   2.2$ & $<  5.3$ & $ $ & $ $ & $ $ & $ $ & $ $ & $ $ \\
155-338 & 5 35 15.51 & -5 23 37.52 & $ $ & $  9.2 \pm   2.4$ & $< 10.6$ & $ $ & $ $ & $  9.3 \pm   4.6$ & $ 11.2 \pm   3.8$ & $  3.5 \pm   0.8$ & $ $ \\
158-323 & 5 35 15.82 & -5 23 22.50 & $ $ & $  9.5 \pm   2.3$ & $<  9.7$ & $ 11.4 \pm   2.0$ & $ 10.6 \pm   2.9$ & $ 11.2 \pm   1.5$ & $ 10.6 \pm   4.8$ & $  7.3 \pm   0.7$ & $ $ \\
158-327 & 5 35 15.79 & -5 23 26.61 & $ $ & $ 12.8 \pm   2.4$ & $ 18.6 \pm   3.2$ & $ $ & $ $ & $  9.2 \pm   1.4$ & $ 13.0 \pm   5.7$ & $  7.5 \pm   1.3$ & $ $ \\
159-350 & 5 35 15.93 & -5 23 49.96 & $ $ & $ 42.7 \pm   3.7$ & $ 11.9 \pm   3.3$ & $ 13.1 \pm   2.0$ & $ $ & $ 10.5 \pm   4.2$ & $ 16.2 \pm   5.7$ & $  7.6 \pm   0.8$ & $ $ \\
163-317 & 5 35 16.27 & -5 23 16.72 & $ 33.3 \pm   3.8$ & $  7.6 \pm   2.0$ & $ 11.3 \pm   2.7$ & $ $ & $ 10.1 \pm   2.5$ & $ 11.1 \pm   1.2$ & $  9.5 \pm   2.8$ & $ 10.8 \pm   2.2$ & $ $ \\
167-317 & 5 35 16.73 & -5 23 16.63 & $ 21.9 \pm   4.1$ & $ 15.0 \pm   2.2$ & $ 19.1 \pm   3.3$ & $ 25.8 \pm   2.0$ & $ 23.3 \pm   4.2$ & $ 25.5 \pm   5.0$ & $ 19.8 \pm   6.0$ & $  6.8 \pm   0.8$ & $ $ \\
168-326NS & 5 35 16.82 & -5 23 26.21 & $ $ & $ 13.6 \pm   2.3$ & $<  7.6$ & $ 10.1 \pm   2.0$ & $ 14.8 \pm   5.0$ & $  8.4 \pm   0.3$ & $ $ & $ $ & $ $ \\
170-337 & 5 35 16.96 & -5 23 37.04 & $ 38.1 \pm   5.2$ & $ 12.9 \pm   2.2$ & $<  5.9$ & $ $ & $ $ & $  6.6 \pm   2.4$ & $ 13.6 \pm   3.3$ & $  6.8 \pm   0.7$ & $ $ \\
171-340 & 5 35 17.05 & -5 23 39.59 & $ 18.3 \pm   4.6$ & $ 13.0 \pm   2.3$ & $<  6.1$ & $ $ & $ $ & $ $ & $ $ & $ $ & $ $ \\
177-341 & 5 35 17.67 & -5 23 40.96 & $ $ & $ 15.8 \pm   2.1$ & $ 16.7 \pm   2.8$ & $ $ & $ $ & $ 10.8 \pm   3.7$ & $ 14.4 \pm   4.0$ & $  7.7 \pm   0.7$ & $ $ \\
HC180 & 5 35 17.39 & -5 24 0.30 & $ $ & $ 10.7 \pm   3.0$ & $<  4.6$ & $ $ & $ $ & $ $ & $ $ & $ $ & $ $ \\
HC189 & 5 35 14.53 & -5 23 56.00 & $ $ & $ 99.6 \pm   8.4$ & $<  9.1$ & $ $ & $ $ & $ $ & $ $ & $ $ & $ $ \\
HC246 & 5 35 15.68 & -5 23 39.10 & $ $ & $ 17.8 \pm   2.4$ & $<  7.5$ & $ $ & $ $ & $ $ & $ $ & $ $ & $ $ \\
HC254 & 5 35 13.86 & -5 23 35.00 & $ $ & $ 17.7 \pm   3.8$ & $< 14.6$ & $ $ & $ $ & $ $ & $ $ & $ $ & $ $ \\
HC295 & 5 35 17.57 & -5 23 24.90 & $ $ & $ 11.1 \pm   2.1$ & $<  5.9$ & $ $ & $ $ & $  4.4 \pm   0.6$ & $  5.9 \pm   2.6$ & $ $ & $ $ \\
HC336 & 5 35 15.81 & -5 23 14.30 & $ $ & $  6.7 \pm   2.0$ & $ 19.7 \pm   3.1$ & $ $ & $ 13.8 \pm   1.0$ & $ 10.0 \pm   5.0$ & $ 10.0 \pm   5.0$ & $  3.7 \pm   0.7$ & $ $ \\
HC350 & 5 35 16.06 & -5 23 7.30 & $ $ & $  8.8 \pm   2.1$ & $<  6.1$ & $ $ & $  3.5 \pm   2.2$ & $  4.1 \pm   0.8$ & $  6.6 \pm   2.0$ & $ $ & $ $ \\
HC351 & 5 35 19.07 & -5 23 7.50 & $ $ & $  8.7 \pm   2.8$ & $<  4.6$ & $ $ & $ $ & $ $ & $ $ & $ $ & $ $ \\
HC361 & 5 35 14.29 & -5 23 4.30 & $ $ & $ 19.2 \pm   3.2$ & $<  7.6$ & $ $ & $ $ & $ $ & $ $ & $ $ & $ $ \\
HC383 & 5 35 17.84 & -5 22 58.20 & $ $ & $  7.0 \pm   2.3$ & $<  5.2$ & $ $ & $ $ & $ $ & $ $ & $ $ & $ $ \\
HC401 & 5 35 16.08 & -5 22 54.10 & $ $ & $  7.3 \pm   2.1$ & $<  5.6$ & $ $ & $ $ & $ $ & $ $ & $ $ & $ $ \\
HC412 & 5 35 16.34 & -5 22 49.10 & $ $ & $  7.7 \pm   2.3$ & $<  5.6$ & $ $ & $ $ & $ $ & $ $ & $ $ & $ $ \\
HC414 & 5 35 16.98 & -5 22 48.50 & $ $ & $  9.1 \pm   2.2$ & $<  5.2$ & $ $ & $ $ & $  3.8 \pm   0.9$ & $  7.0 \pm   5.6$ & $ 10.5 \pm   0.5$ & $ $ \\
HC418 & 5 35 18.08 & -5 22 47.10 & $ $ & $  7.6 \pm   2.5$ & $<  5.5$ & $ $ & $ $ & $ $ & $ $ & $ $ & $ $ \\
HC436 & 5 35 18.38 & -5 22 37.50 & $ $ & $  8.7 \pm   2.8$ & $<  5.5$ & $ $ & $ 16.6 \pm   1.3$ & $  9.0 \pm   0.3$ & $ $ & $ 11.2 \pm   0.4$ & $ $ \\
HC438 & 5 35 14.09 & -5 22 36.60 & $ $ & $ 67.8 \pm  14.2$ & $< 12.1$ & $ $ & $ $ & $  2.0 \pm   0.2$ & $ $ & $ $ & $ $ \\
HC440 & 5 35 17.36 & -5 22 35.80 & $ $ & $ 13.0 \pm   2.8$ & $<  7.0$ & $ $ & $ $ & $ $ & $ $ & $ $ & $ $ \\
HC495 & 5 35 13.52 & -5 22 19.60 & $ $ & $ 22.1 \pm   6.6$ & $<  6.4$ & $ $ & $ $ & $ $ & $ $ & $ $ & $ $ \\
HC498 & 5 35 18.96 & -5 22 18.80 & $ $ & $ 15.3 \pm   4.6$ & $<  4.3$ & $ $ & $ $ & $ $ & $ $ & $ $ & $ $ \\
HC514 & 5 35 16.43 & -5 22 12.20 & $ $ & $ 23.7 \pm   7.4$ & $<  4.6$ & $ $ & $ $ & $ $ & $ $ & $ $ & $ $ \\
HC771 & 5 35 14.86 & -5 22 44.10 & $ $ & $ 16.1 \pm   4.9$ & $<  6.7$ & $ $ & $ $ & $ $ & $ $ & $ $ & $ $ \\
LMLA162 & 5 35 14.40 & -5 23 50.84 & $ $ & $103.3 \pm   7.2$ & $< 10.0$ & $ $ & $  1.0 \pm   0.1$ & $ $ & $ $ & $ $ & $ $ \\
MM8 & 5 35 13.73 & -5 23 46.84 & $ $ & $407.2 \pm  27.5$ & $ 28.8 \pm   5.9$ & $ $ & $  0.8 \pm   0.1$ & $ $ & $ $ & $ $ & $ $ \\
MM13 & 5 35 13.75 & -5 24 7.74 & $ $ & $317.5 \pm  25.1$ & $ 36.9 \pm   8.3$ & $ $ & $ $ & $ $ & $ $ & $ $ & $ $ \\
MM21 & 5 35 13.57 & -5 23 59.04 & $ $ & $153.6 \pm  13.2$ & $ 15.6 \pm   4.7$ & $ $ & $  0.9 \pm   0.1$ & $ $ & $ $ & $ $ & $ $ \\
MM22 & 5 35 13.66 & -5 23 54.94 & $ $ & $123.0 \pm  12.7$ & $ 17.1 \pm   4.2$ & $ $ & $  0.4 \pm   0.1$ & $ $ & $ $ & $ $ & $ $ \\
MM23 & 5 35 14.00 & -5 22 45.04 & $ $ & $ 61.2 \pm   7.4$ & $  9.6 \pm   3.1$ & $ $ & $ $ & $ $ & $ $ & $ $ & $ $ \\
MM24 & 5 35 14.62 & -5 22 28.94 & $ $ & $167.4 \pm  27.0$ & $< 32.7$ & $ $ & $ $ & $ $ & $ $ & $ $ & $ $ \\
\enddata
\tablecomments{
All quoted uncertainties are
1$\sigma$; uncertainties (of 10\%) in the overall flux scale
are not included in this table.  The MM sources are detected only at
wavelengths $\ga 1$ mm.  HC objects are known near-IR cluster members
from \citet{HC00} that are also detected at mm wavelengths, and the sources
with numerical labels are a subset of this sample that are also detected
optically as proplyds \citep[e.g.,][]{OWH93}.  
LMLA162 is absent from \citet{HC00}, 
but seen at 3.6 $\mu$m by \citet{LADA+04}; we also see a faint 2 $\mu$m point
source at this location in a 2MASS image. MM8 and MM13
were detected by \citet{EC06}, and MM21 was detected by \citet{EC06}
but associated with HC 178 (we find that MM21 and HC 178 are separated
by an angle larger than our relative positional uncertainties).
LMLA162, MM8, MM13, MM21, and MM22 were also detected in previous
1.3 mm observations by \citet{ZAPATA+05}.  MM23 and MM24 are newly detected 
here.}
\tablerefs{References for the fluxes are as follows: 880 $\mu$m
\citep{WAW05}; 1.3 mm (this work); 3 mm \citep{EC06}; 3.6 mm \citep{MLL95};
1.3 cm \citep{ZAPATA+04}; 2 and 6 cm \citep{FELLI+93b}; 20 cm 
\citep{FELLI+93a}.  Quoted upper limits for the 3 mm fluxes are 3$\sigma$.}
\end{deluxetable}

\begin{deluxetable}{lcccc}
\tabletypesize{\scriptsize}
\tablewidth{0pt}
\tablecaption{Derived quantities for detected sources
\label{tab:results}}
\tablehead{\colhead{ID} & {$S_{\rm \nu, dust}$ (mJy)} & 
\colhead{$M_{\rm circumstellar}$ (M$_{\odot}$)} & 
\colhead{$R_{\rm disk}$ (AU)} & \colhead{$S_{\nu, \tau \ga 1}$ (mJy)}} 
\startdata
147-323 & $  7.5 \pm   2.2$ &  0.009 $\pm$  0.003 & 88 &    83 \\
155-338 & $  4.0 \pm   4.0$ &  0.005 $\pm$  0.005 & 102 &   112 \\
158-323 & $  1.0 \pm   1.0$ &  0.001 $\pm$  0.001 & 105 &   119 \\
158-327 & $  5.0 \pm   4.0$ &  0.006 $\pm$  0.005 & 122 &   161 \\
159-350 & $ 33.0 \pm   7.0$ &  0.042 $\pm$  0.009 & 152 &   250 \\
163-317 & $  6.0 \pm   2.0$ &  0.008 $\pm$  0.003 & 93 &    93 \\
167-317 & $  2.0 \pm   2.0$ &  0.003 $\pm$  0.003 & 122 &   161 \\
168-326NS & $  7.0 \pm   4.0$ &  0.009 $\pm$  0.005 & $<$100 & $<108$ \\
170-337 & $ 10.0 \pm   2.5$ &  0.013 $\pm$  0.003 & 126 &   171 \\
171-340 & $ 13.0 \pm   2.3$ &  0.016 $\pm$  0.003 & 80 &    69 \\
177-341 & $  4.0 \pm   4.0$ &  0.005 $\pm$  0.005 & 177 &   339 \\
HC180 & $ 10.7 \pm   3.0$ &  0.013 $\pm$  0.004 & $<$100 & $<108$ \\
HC189 & $ 99.5 \pm   8.4$ &  0.125 $\pm$  0.011 & $<$100 & $<108$ \\
HC246 & $ 17.8 \pm   2.4$ &  0.022 $\pm$  0.003 & $<$100 & $<108$ \\
HC254 & $ 17.7 \pm   3.8$ &  0.022 $\pm$  0.005 & $<$100 & $<108$ \\
HC295 & $  8.0 \pm   3.5$ &  0.010 $\pm$  0.004 & 187 &   381 \\
HC336 & $  0.0 \pm   0.0$ &  0.000 $\pm$  0.000 & $<$100 & $<108$ \\
HC350 & $  5.0 \pm   3.5$ &  0.006 $\pm$  0.004 & $<$100 & $<108$ \\
HC351 & $  8.7 \pm   2.8$ &  0.011 $\pm$  0.003 & $<$100 & $<108$ \\
HC361 & $ 19.2 \pm   3.2$ &  0.024 $\pm$  0.004 & $<$100 & $<108$ \\
HC383 & $  7.0 \pm   2.3$ &  0.009 $\pm$  0.003 & 202 &   441 \\
HC401 & $  7.3 \pm   2.1$ &  0.009 $\pm$  0.003 & $<$100 & $<108$ \\
HC412 & $  7.7 \pm   2.3$ &  0.010 $\pm$  0.003 & 161 &   283 \\
HC414 & $  3.0 \pm   3.0$ &  0.004 $\pm$  0.004 & $<$100 & $<108$ \\
HC418 & $  7.6 \pm   2.5$ &  0.010 $\pm$  0.003 & $<$100 & $<108$ \\
HC436 & $  1.0 \pm   1.0$ &  0.001 $\pm$  0.001 & $<$100 & $<108$ \\
HC438 & $ 67.8 \pm  14.2$ &  0.085 $\pm$  0.018 & 200 &   433 \\
HC440 & $ 13.0 \pm   2.8$ &  0.016 $\pm$  0.004 & $<$100 & $<108$ \\
HC495 & $ 22.1 \pm   6.6$ &  0.028 $\pm$  0.008 & $<$100 & $<108$ \\
HC498 & $ 15.3 \pm   4.6$ &  0.019 $\pm$  0.006 & $<$100 & $<108$ \\
HC514 & $ 23.7 \pm   7.4$ &  0.030 $\pm$  0.009 & 166 &   299 \\
HC771 & $ 16.1 \pm   4.9$ &  0.020 $\pm$  0.006 & 176 &   337 \\
LMLA162 & $103.3 \pm   7.2$ &  0.130 $\pm$  0.009 & $<$100 & $<108$ \\
MM8 & $407.2 \pm  27.5$ &  0.512 $\pm$  0.035 & $<$100 & $<108$ \\
MM13 & $317.5 \pm  25.1$ &  0.399 $\pm$  0.032 & $<$100 & $<108$ \\
MM21 & $153.6 \pm  13.2$ &  0.193 $\pm$  0.017 & $<$100 & $<108$ \\
MM22 & $123.0 \pm  12.7$ &  0.155 $\pm$  0.016 & $<$100 & $<108$ \\
MM23 & $ 61.2 \pm   7.4$ &  0.077 $\pm$  0.009 & $<$100 & $<108$ \\
MM24 & $167.3 \pm  27.0$ &  0.211 $\pm$  0.034 & 217 &   514 \\
\enddata
\tablecomments{$S_{\rm \nu,dust}$ is the component of the observed
1.3 mm emission due to cool dust, determined from a fit to long-wavelength
fluxes of a model including thermal free-free emission as well as dust 
emission (see Figure \ref{fig:seds}).  Quoted uncertainties are
1$\sigma$, and do not include systematic uncertainties associated with the 
overall flux calibration or conversion from flux to mass.}
\end{deluxetable}


\epsscale{1.0}

\epsscale{1.0}

\epsscale{1.0}

\epsscale{1.0}


\epsscale{1.0}

\clearpage
\epsscale{1.0}
{\plotone{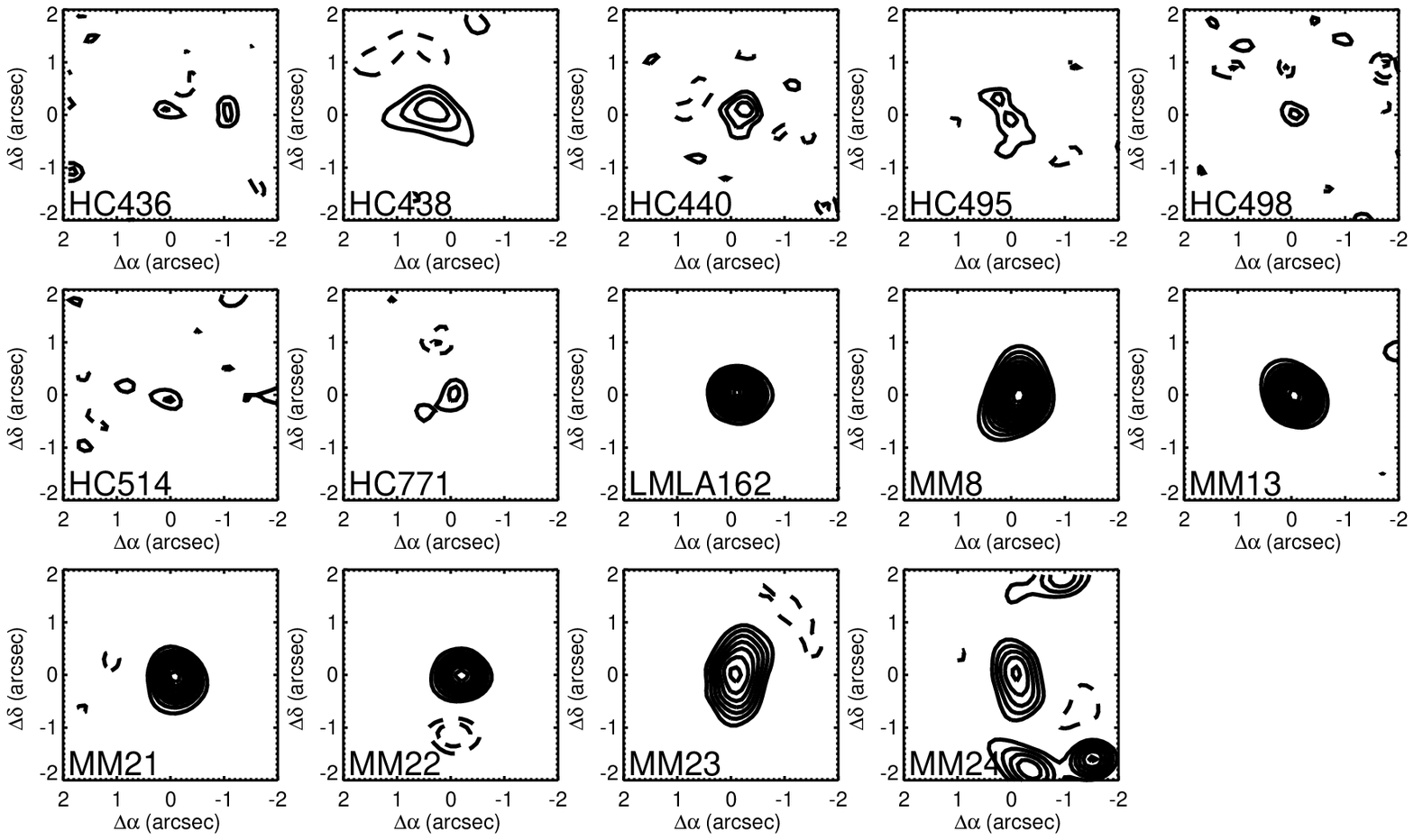}}\\[5mm]
\centerline{Fig. 5. --- continued.}
\clearpage

\epsscale{1.0}

\clearpage
\epsscale{1.0}
{\plotone{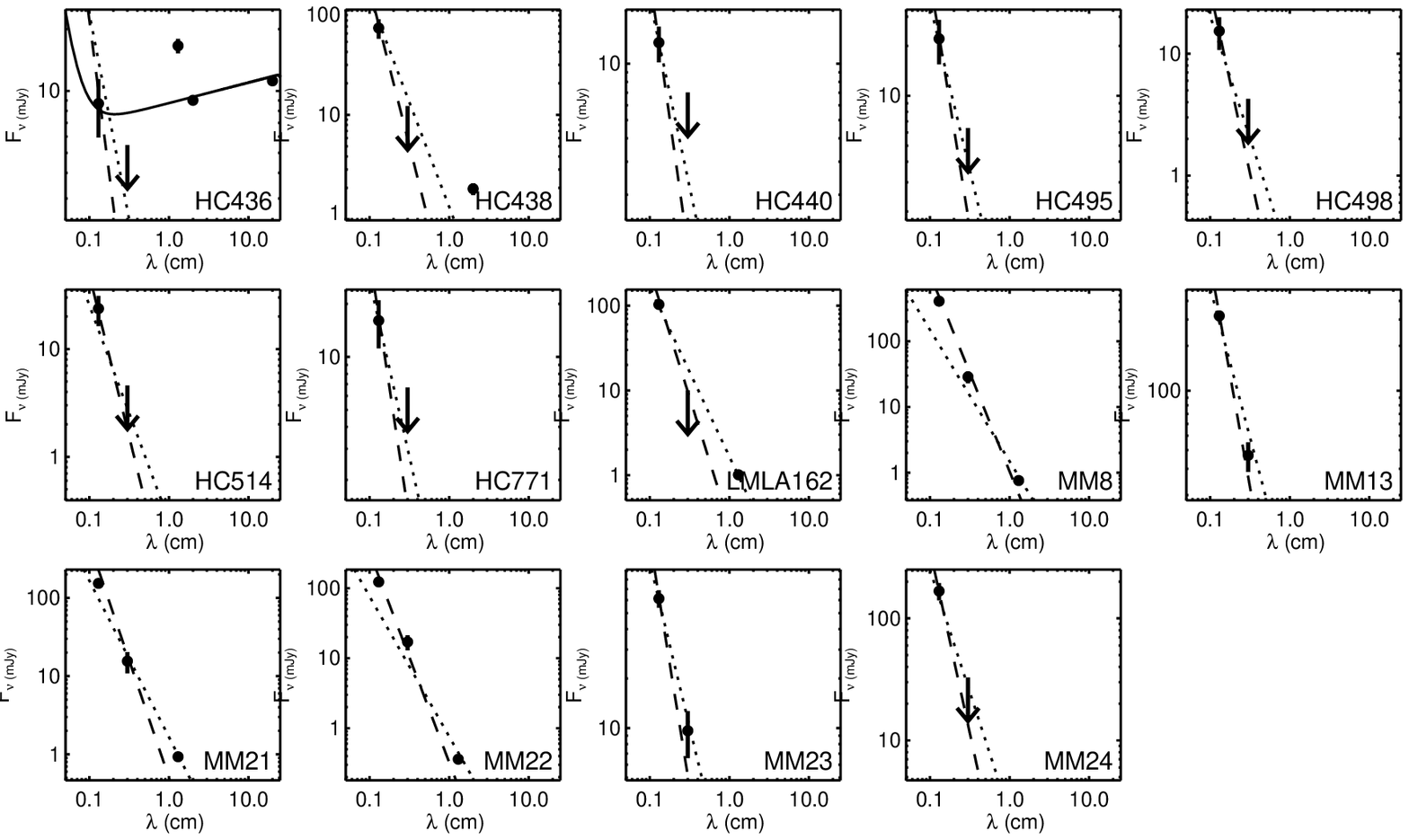}}\\[5mm]
\centerline{Fig. 6. --- continued.}
\clearpage

\epsscale{1.0}

\epsscale{1.0}

\epsscale{1.0}


\epsscale{1.0}

\epsscale{1.0}

\end{document}